\documentclass[prd,eqsecnum,floatfix]{revtex4}%
\usepackage{amsmath}
\usepackage{graphicx}
\usepackage{amsfonts}
\usepackage{amssymb}%
\providecommand{\U}[1]{\protect\rule{.1in}{.1in}}
\newcommand{\eqn}[1]{(\ref{#1})}

\newcommand{\BOX}{\hbox {$\sqcap$ \kern -1em $\sqcup$}}

\newcommand{\be}{\begin{equation}}
\newcommand{\ee}{\end{equation}}
\newcommand{\ba}{\begin{eqnarray}}
\newcommand{\ea}{\end{eqnarray}}
\newcommand{\ban}{\begin{eqnarray*}}
\newcommand{\bea}{\begin{eqnarray}}
\newcommand{\eea}{\end{eqnarray}}
\newcommand{\ean}{\end{eqnarray*}}
\newcommand{\barr}{\begin{array}}
\newcommand{\earr}{\end{array}}

\newcommand{\nn}{\nonumber}
\newcommand{\tr}{\mbox{tr}}
\def\e{{\rm e}}

\begin{document}
\title{Non-extensive statistics in stringy space-time foam models and entangled meson states}
\author{Nick E. Mavromatos and Sarben Sarkar}
\affiliation{King's College London, University of London, Department of Physics, Strand,
London WC2R 2LS, U.K.}

\begin{abstract}

The possibility of generation of non-extensive statistics, in the sense of
Tsallis, due to space-time foam is discussed within the context of a
particular kind of foam in string/brane-theory, the D-particle foam model .
The latter involves point-like brane defects (D-particles), which provide the
topologically non-trivial foamy structures of space-time. A stochastic
Langevin equation for the velocity recoil of D-particles can be derived from the pinched approximation for a sum over genera in the calculation of the partition
function of a bosonic string in the presence of heavy D-particles. The string coupling
in standard perturbation theory is related to the exponential of the
expectation of the dilaton. Inclusion of fluctuations of the dilaton itself and uncertainties in the string background
will then necessitate fluctuations in $g_{s}$. The
fluctuation in the string coupling in the sum over genera typically leads to a generic structure of the Langevin equation where the coefficient of the noise term fluctuates owing to dependence on the string coupling $g_{s}$.  The positivity of $g_{s}$ leads
naturally to a stochastic modelling of its distribution with a $\chi$-
distribution. This then rigorously implies a Tsallis type non-extensive or, more generally, a superstatistics
distribution for the recoil velocity of D-particles. As a concrete and physically interesting
application, we provide a rigorous estimate of an $\mathbb{\omega}$-like effect, pertinent to CPT violating modifications of the Einstein-Podolsky-Rosen
correlators in entangled states of  neutral Kaons. In the case of D-particle foam fluctuations, which respect the Lorentz symmetry of the vacuum on average, we find that the $\mathbb{\omega}$-effect may be within the range of sensitivity of future meson factories.

\end{abstract}
\maketitle

%\date{\today}

\section{Introduction}

The standard model of particle physics is considered to be successful even
though it has many undetermined parameters which need to be fitted to data.
One reason for the great interest in physics beyond the standard model (SM) is
the absence, within its framework, of quantum gravity (QG).
A\ full picture of QG still
remains elusive. One interesting approach to it, based on string theory (ST)
\cite{green,polch}, has the advantage of \ also unifying QG with the nuclear and
electromagnetic forces.
Moreover flavour
mixing phenomena are not understood in the sense that phenomenological mixing
matrices are not derived from a fundamental point of view. There may be some
relationship between these two deficiencies, and QG may play a r\^ole in that.
For instance, (a small) part of the observed \emph{mass differences} between neutrinos
might be quantum gravitational in origin
as argued recently~\cite{sabmav}
within the context of a space-time foam model in string theory with space-time defects~\cite{Dfoam} in the form of
point-like D(irichlet)-branes (D-particles)~\cite{polch,polch2,johnson}. The interaction of the defects with string matter in such a model induces flavour mixing and non-trivial contributions to the cosmological constant or better vacuum energy) proportional to the mixing angle and the mass differences among the neutrino states.

Because of its foundations based on local
relativistic unitary quantum field theory, SM has CPT symmetry \cite{streater}.
Hence the detection of possible violations of CPT necessitates clearly physics beyond the standard
model. CPT violation might be an important feature of QG, which may be exhibited by some stringy models of space-time foam with defects, of the type discussed in \cite{Dfoam}.
Recently, an interesting signature of CPT Violation due to decoherence of matter as a result of QG effects
in stochastic space-time foam models,
has been suggested in ref.~\cite{bernabeu1},
implying new physics beyond SM  in the entanglement properties of neutral
meson pairs. The proposed signature is the so called omega effect
where modifications to the standard Einstein-Podolsky-Rosen (EPR) correlations appear for entangled states of neutral flavoured mesons in  meson factories~\cite{bernabeu1,bernabeu}.

Neutral mesons, such as the $K$ mesons, have in the past been pivotal in the study of
discrete symmetries \cite{wu}. The decay of a (generic) meson (such as. the
$\phi$ meson produced in \ collisions of $e^{+}$ and $e^{-}$ with quantum
numbers $J^{PC}=1^{--}$ \cite{lipkin}), leads to a pair state $\left\vert
i\right\rangle $ of neutral mesons ($M$) which has the form of the entangled
state%
\begin{equation}
\left\vert i\right\rangle =\frac{1}{\sqrt{2}}\left(  \left\vert \overline
{M_{0}}\left(  \overrightarrow{k}\right)  \right\rangle \left\vert
M_{0}\left(  -\overrightarrow{k}\right)  \right\rangle -\left\vert
M_{0}\left(  \overrightarrow{k}\right)  \right\rangle \left\vert
\overline{M_{0}}\left(  -\overrightarrow{k}\right)  \right\rangle \right)  .
\end{equation}
This state has $CP=+$. If CPT is not well-defined (actually \emph{perturbatively}, i.e. although
the concept of the anti-particle still exists,it is slightly modified),
then $M_{0}$ and $\overline
{M_{0}}$ may not be identified and the requirement of $CP=+$ can be relaxed
\cite{bernabeu1},\cite{bernabeu}. Consequently the state of the meson pair can
be parametrised to have the form
\begin{align}
\left\vert i\right\rangle  &  =\left(  \left\vert \overline{M_{0}}\left(
\overrightarrow{k}\right)  \right\rangle \left\vert M_{0}\left(
-\overrightarrow{k}\right)  \right\rangle -\left\vert M_{0}\left(
\overrightarrow{k}\right)  \right\rangle \left\vert \overline{M_{0}}\left(
-\overrightarrow{k}\right)  \right\rangle \right) \nonumber\\
&  +\mathbb{\omega}\left(  \left\vert \overline{M_{0}}\left(  \overrightarrow
{k}\right)  \right\rangle \left\vert M_{0}\left(  -\overrightarrow{k}\right)
\right\rangle +\left\vert M_{0}\left(  \overrightarrow{k}\right)
\right\rangle \left\vert \overline{M_{0}}\left(  -\overrightarrow{k}\right)
\right\rangle \right)  \label{omega}%
\end{align}
where $\mathbb{\omega}=\left\vert\mathbb{\omega}\right\vert \e^{i\Omega}$ is a complex CPT
violating (CPTV) parameter \cite{bernabeu1}. It turns out to be difficult to
generate this $\mathbb{\omega}$ term since reasonable attempts based on local effective
lagrangians with explicit CPT violating terms do not lead to it. Recent
approaches have concentrated on properties of space-time foam and related
decoherence, which by their nature go beyond the concept of local effective field theory.
From extrapolations of black hole physics and quantum uncertainty
relations to the smallest scales it seems unavoidable that space and time
should fluctuate on a space scale $l_{P}$, the Planck length, and a time scale
$t_{P}$, the Planck time. Such quantum fluctuations lead to a background state
which will be generically called space-time foam \cite{wheeler}. The full
understanding of quantum states of matter in such a medium foam requires a
deeper understanding of QG. Consequently the approaches to it have ranged from
the purely phenomenological to those based on a more fundamental theory such
as strings. Even in the latter, phenomenological assumptions will have to be
made since we do not have a developed string field theory and formulations
have typically been based on fixed backgrounds. Indeed a plausible
phenomenological approach based on thermal baths, at least in its natural and
most straightforward form, cannot generate an $\mathbb{\omega}$ term \cite{sarkar}. On
the other hand a stochastic approach based and inspired by theoretical
considerations of D-particle recoil from stringy matter in D-particle foam models~\cite{Dfoam} has predicted and quantified this omega effect \cite{bernabeu}. However, the quantification was based on somewhat phenomenological and
naive estimates. In this work we shall provide more rigorous estimates of this effect within D-particle foam models.

In the treatments so far of D-particle foam the stochastic modelling has
concentrated on the randomness of the recoil velocity vector of the D-particle
defect during its topologically non-trivial interaction with the matter string state (i.e. the capture and subsequent re-emission of the matter state). In this work we
want to incorporate the effects of the (target-space) quantum fluctuations of the D-particle
which, within the bosonic string
model, can be calculated using perturbation theory. As we shall discuss, the renormalization of a certain subleading divergence to all orders in
string perturbation theory leads to fluctuations superposed on a drift
velocity of the D-particle obtained after the initial recoil on interacting
with stringy matter. Furthermore, at the phenomenological level uncertainty in
the string vacua can lead to a fluctuating string coupling $g_{s}$ which is
related to \ vacuum expectation value of the dilaton field. Incorporation of
such fluctuations lead to superstatistics \cite{beck2} of the velocity recoil
distribution, non-extensive statistics of Tsallis \cite{tsallis} being one
possibility. The recoil fluctuations of the D-particle lead to effective
stochastic back-reaction on space-time that cannot be neglected. Hence the
space-time metric will have induced stochastic contributions from
stochasticity in the recoil \cite{mavromatos}.

In this work we shall concentrate our
discussion on the Bosonic string. Although admittedly, this is not relevant phenomenologically,
it is the only case where we manage to sum over world sheet genera and express our basic results on stochastic
fluctuations of space-time and fuzziness of string coupling in a closed form. We shall make some comments on the robustness of our results and thus their extension to supersymmetric cases at the end of our article. This constitutes an active part of our research at present.

 Within the Bosonic $\sigma$-model framework of D-particle foam, we shall calculate the stochastic fluctuations of the space-time
induced by D-particle recoil, and then we shall use it to estimate the
resulting $\mathbb{\omega}$-effect in the initial state of entangled mesons in this concrete microscopic model.
Contrary to previous naive estimates~\cite{bernabeu}, based on dimensional analysis, in this particular example of (string-inspired) foam, the effect is calculated 
rather rigorously, following conformal field theory methods in the world-sheet of the string. In particular, for the magnitude of the $\mathbb{\omega}$-effect in the initial entangled meson states (\ref{omega}), we find the following result for the square of the amplitude of the parameter $\mathbb{\omega}$:
\begin{equation}
|\omega|^2 \sim \frac{m_1^2 + m_2^2}{(m_1 - m_2)^2}\frac{k^2}{M_P^2}~,
\label{omegasup}
\end{equation}
for non relativistic entangled states of mesons, with (near degenerate) masses $m_i~, i=1,2$ and $M_P$ the quantum gravity scale, assumed to be the four dimensional Planck scale, $10^{19}$ GeV.
For comparison, we remind the reader that the naive phenomenological estimates of \cite{bernabeu}
lead to effects of order $|\mathbb{\omega}|^2_{\rm naive} \sim \zeta^2 \frac{k^4}{M_P^2(m_1 - m_2)^2}~$, with $\zeta$ a typical momentum transfer variable, during the interaction of the meson state with the
D-particle, $\zeta \sim \frac{\Delta {\vec k}}{k}$, which in \cite{bernabeu} has been assumed not very small. What the detailed string calculation in this paper shows, in which the stochastic space-time fluctuations are due to quantum fluctuations of recoil-velocities about a zero (Lorentz invariant) average, is that the parameter 
$\zeta$ of \cite{bernabeu} depends on details of the meson system, such as the mass and momenta of the mass eigenstates, and is effectively of order $\zeta^2 \sim (m_1^2 + m_2^2)/k^2$~.
For the case of neutral kaons in $\phi$-factories, with $k \sim {\cal O}(1 ~{\rm GeV})$,
this is already at the order of magnitude required for detectability in an upgrade in DA$\Phi$NE~\cite{bernabeu}. Hence these models of D-particle foam may be falsifiable in such future meson factories.
Of course, the estimate assumes the neutral Kaon as interacting with the space-time foam as an elementary entity, ignoring details of strongly interacting constituents in the Kaon substructure. Taking into account such strong interaction effects might change the above estimates.

The structure of the article is as follows: in the next section we will discuss general features of
the D-particle foam model, on which our work is based. In section \ref{sec:3}
we review the basic mathematical properties of the deformations of the bosonic sigma-model, describing the interaction of string matter with  D-particle defects. In section \ref{sec:4} we discuss re-summation of higher world-sheet topologies and their implication on inducing stochastic fluctuations in the recoil velocity of the D-particle. The latter are expressed through a Langevin-type stochastic equation, which replaces the tree-level world-sheet renormalization group equation for the recoil velocity, interpreted as a
sigma-model coupling. In section \ref{sec:5}, we solve this Langevin equation but also discuss ``fuzzy'' aspects of the string coupling in a full quantum-gravity setting of this string model of foam. For D-particle foam we thus arrive, in the sense of Tsallis, at a superstatistics description.
We discuss the application of our results to a precise estimation of decoherence-induced CPT-violating effects in entangled states of mesons (the $\mathbb{\omega}$-effect) in section \ref{sec:6}, where we estimate the effect in this model of (D-particle) space-time and compare it to previous estimates.
Our conclusions and outlook are presented in section \ref{sec:7}. Some technical aspects  are reviewed in two Appendices.

\section{Generic Features of D-particle Foam Model \label{sec:2}}

\begin{figure}[t]
\centering
\includegraphics[width=7.5cm]{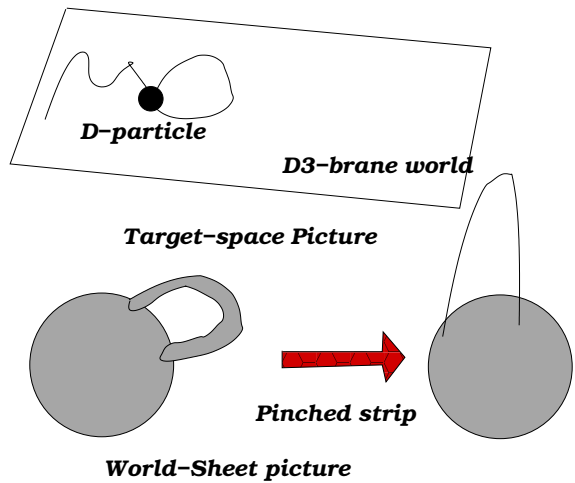}\caption{\emph{Upper picture:} A
fluctuating D-particle is described by open strings attached to it. As a
result of conservation of  string fluxes~\cite{polch,polch2,johnson} that accompany
the D-branes, an isolated D-particle cannot occur, but it has to be connected
to a D-brane world through flux strings. \emph{Lower picture}: World-sheet
diagrams with annulus topologies, describing the fluctuations of D-particles
as a result of the open string states ending on them. Conformal invariance
implies that pinched surfaces, with infinitely long thin strips, have to be
taken into account. In bosonic string theory, such surfaces can be
resummed~\cite{szabo}. }%
\label{fig:dbranes}%
\end{figure}

\begin{figure}[th]
\centering
\includegraphics[width=7.5cm]{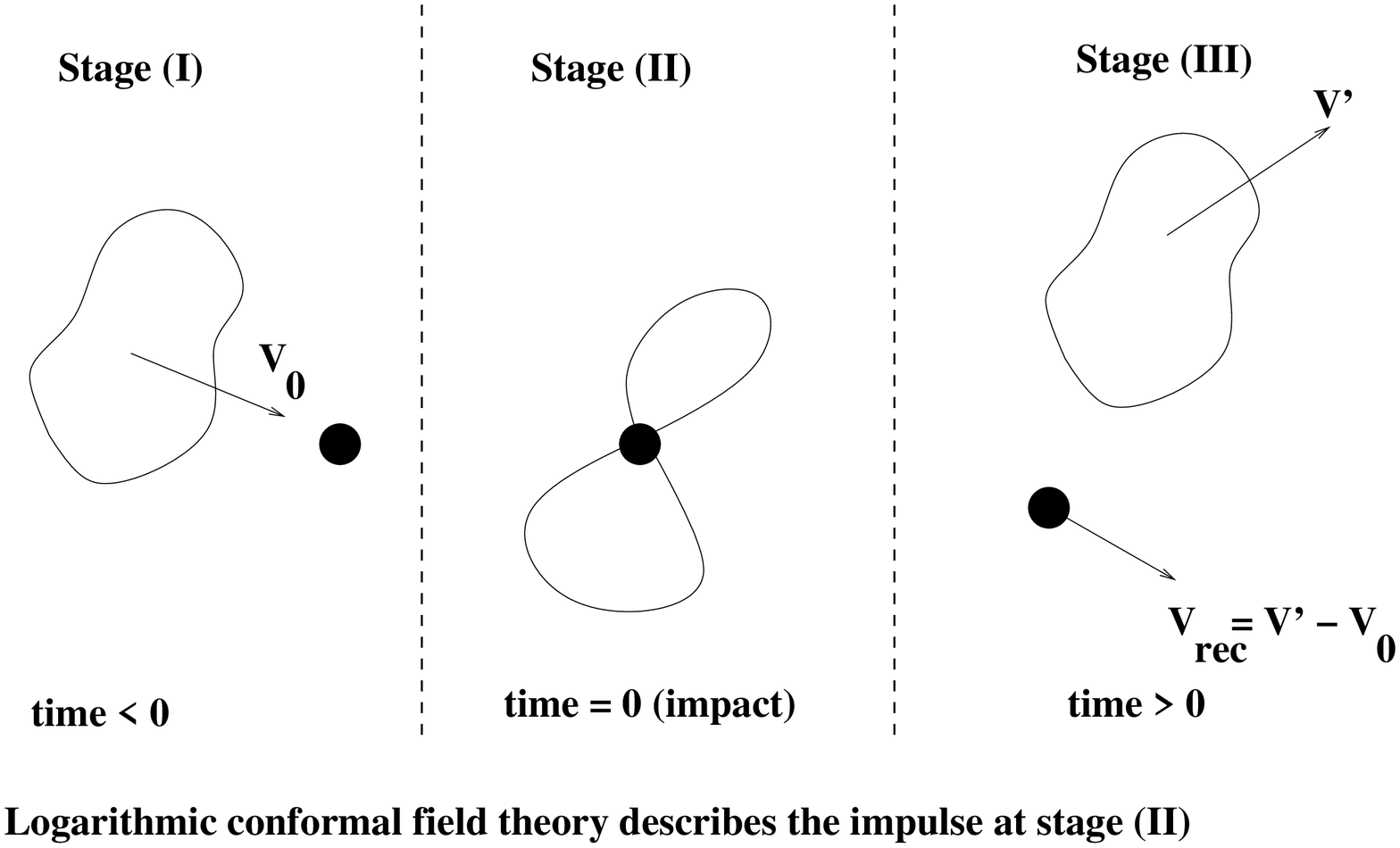} \hfill
\includegraphics[width=7.5cm]{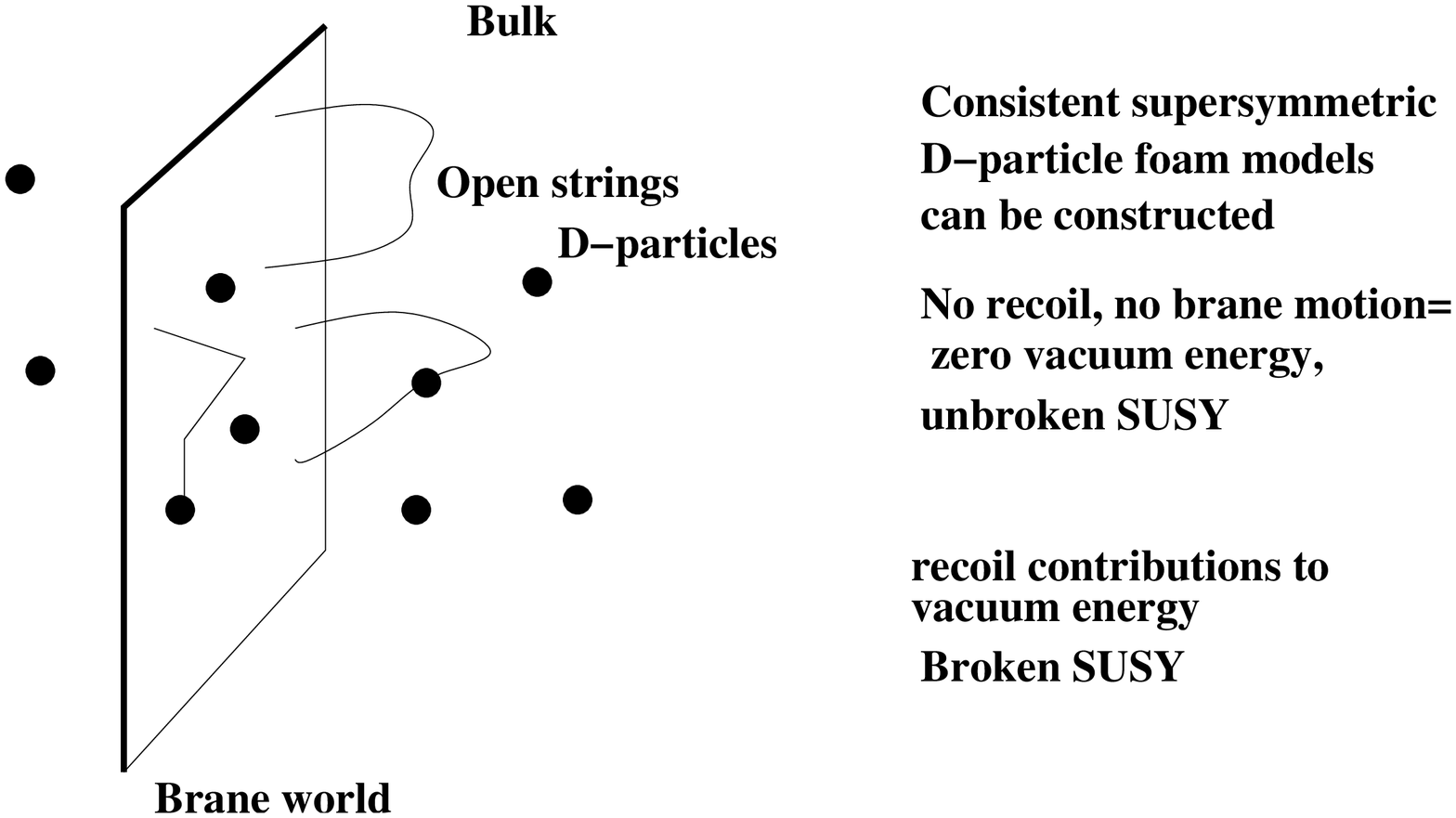} \caption{Schematic
representation of a D-foam. The figure indicates also the capture/recoil
process of a string state by a D-particle defect for closed (left) and open
(right) string states, in the presence of D-brane world. The presence of a
D-brane is essential due to gauge flux conservation, since an isolated
D-particle cannot exist. The intermediate composite state at $t=0$, which has
a life time within the stringy uncertainty time interval $\delta t$, of the
order of the string length, and is described by world-sheet logarithmic
conformal field theory, is responsible for the distortion of the surrounding
space time during the scattering, and subsequently leads to induced metrics
depending on both coordinates and momenta of the string state. This results on
modified dispersion relations for the open string propagation in such a
situation~\cite{Dfoam}, leading to non-trivial \textquotedblleft
optics\textquotedblright\ for this space time.}%
\label{fig:recoil}%
\end{figure}

In this section we shall review briefly the basic features of the D-particle
foam model, discussed in \cite{Dfoam}.
We will use some established results and constructs from string/brane theory
\cite{polch,polch2}, which we shall discuss briefly for the benefit of the non-expert reader.
In particular, zero dimensional D-branes \cite{johnson}
occur (in bosonic and some supersymmetric string theories) and are also known
as D-particles. Interactions in string theory are, as yet, not treated as
systematically as in ordinary quantum field theory where a second quantised
formalism is defined. The latter leads to the standard formulations by
Schwinger and Feynman of perturbation series. When we consider stringy matter
interacting with other matter or D-particles, the world lines traced out by
point particles are replaced by two-dimensional world sheets. World sheets are
the parameter space of the first quantised operators ( fermionic or bosonic)
representing strings. In this way the first quantised string is represented by
actually a two dimensional (world-sheet) quantum field theory. An important
consistency requirement of this first quantised string theory is conformal
invariance which determines the space-time dimension and/or structure. \ This
symmetry permits the representation of interactions through the construction
of measures on inequivalent Riemann surfaces \cite{green}. In and out states
of stringy matter are represented by vertex insertions at the boundaries. The
D-particles as solitonic states~\cite{polch2} in string theory do fluctuate
themselves quantum mechanically;  this is described by stringy excitations, corresponding to
open strings with their ends attached to the D-particles and higher
dimensional D branes. In a first quantised (world-sheet) language, such
fluctuations are also described by Riemann surfaces of higher topology with
appropriate Dirichlet boundary conditions (c.f. fig.~\ref{fig:dbranes}). The
plethora of Feynman diagrams in higher order quantum field theory is replaced
by a small set of world sheet diagrams classified by moduli which need to be
summed or integrated over \cite{zwiebach}. The model of space-time foam, used in understanding the omega
effect, is based on D-particles populating a bulk geometry between parallel
D-brane worlds. The model is termed D-foam~\cite{Dfoam} (c.f. figure
\ref{fig:recoil}), and our world is modelled as a three-brane moving in the
bulk geometry; as a result, D-particles cross the brane world and appear for
an observer on the brane as foamy structures which flash on and off .

Even at low energies $E$, such a foam may have observable consequences e.g.
decoherence \ effects which may be of magnitude $O\left(  \left[  \frac
{E}{M_{P}}\right]  ^{n}\right)  $ with $n=1,2$ where $M_{P}$ is the Planck
mass or change in the usual Lorentz invariant dispersion relations.The study
of D-brane dynamics has been made possible by Polchinski's realisation~\cite{polch2} that
such solitonic string backgrounds can be described in a conformally invariant
way in terms of world sheets with boundaries \cite{polch2}. On these
boundaries Dirichlet boundary conditions for the collective target-space
coordinates of the soliton are imposed \cite{coll}. When low energy matter
given by a closed string propagating in a $\left(  d+1\right)  $-dimensional
space-time collides with a very massive D-particle (0-brane) embedded in this
space-time, the D-particle recoils as a result \cite{kogan} in a
non-relativistic manner. We shall consider the simple case of bosonic stringy
matter coupling to D-particles. Hence we can only discuss matters of principle
and ignore issues of stability due to tachyons. However we should note that an
open string model needs to incorporate for completeness, higher dimensional
D-branes such as the D3 brane. This is due to the vectorial charge carried by
the string owing to the Kalb-Ramond field. Higher dimensional D-branes (unlike
D-particles) can carry the charge from the endpoints of open strings that are
attached to them. For a closed bosonic string model the inclusion of such
D-branes is not imperative (see figure \ref{fig:recoil}) although D-particle
fluctuations would generally require them. The details of the higher
dimensional branes are not essential for our analysis however. The current
state of phenomenolgical modelling of the interactions of D-particle foam with
stringy matter will be briefly summarised now. Since there are no rigid bodies
in general relativity the recoil fluctuations of the brane and the effective
stochastic back-reaction on space-time cannot be neglected. As we will
discuss, D-particle recoil in the "tree approximation" i.e. in lowest order in
the string coupling $g_{s}$, is required to cancel infrared singularities in a
higher order disc or Riemann sphere amplitude in open or closed string theory
respectively; the recoil induces a non-trivial space-time metric. For
$\varepsilon$ a positive infinitesimal,  $\Theta_{\varepsilon
}\left(  t\right)  $, \ the regularised step function is introduced, in terms of a contour
integral%
\begin{equation}
\Theta_{\varepsilon}\left(  t\right)  =\frac{1}{2\pi i}\int_{-\infty}^{\infty
}\frac{d\omega}{\omega-i\varepsilon}\e^{i\omega t}.
\end{equation}
For closed strings colliding with a heavy (i.e. non-relativistic) D-particle,
the metric has the form \cite{mav2}
\begin{equation}
g_{ij}=\delta_{ij},\,g_{00}=-1,g_{0i}=\varepsilon\left(  \varepsilon
y_{i}+u_{i}t\right)  \Theta_{\varepsilon}\left(  t\right)  ,\;i=1,\ldots,d.
\end{equation}
where the suffix $0$ denotes temporal (Liouville) components, $u_{i}=\left(
k_{1}-k_{2}\right)  _{i}\;$is small, $k_{1}\left(  k_{2}\right)  $ is the
momentum of the propagating closed-string state before (after) the recoil,
$y_{i}$ are the spatial collective coordinates of the D- particle and
$\varepsilon^{-2}$ is identified with the target Minkowski time $t$ for
$t\gg0$ after the collision. The latter requirement is consistent with
$\varepsilon$ being infinitesimal. For our purposes the Liouville and
Minkowski times can be identified. Now for large $t,$ to leading order,%
\begin{equation}\label{finsler}
g_{0i}\simeq\overline{u}_{i}\equiv\frac{u_{i}}{\varepsilon}\propto g_s\frac{\Delta
p_{i}}{M_{s}}%
\end{equation}
where $\Delta p_{i}$ is the momentum transfer during a collision and $M_s$ is the string mass
scale, $g_s < 1 $ is the string coupling, assumed weak, and the combination
$M_{s}/g_s$
is the D-particle mass, playing the r\^ole of the Quantum Gravity scale in this problem, i.e. the Planck mass; this formalism was used to establish a phenomenological
model where the couplings $u_{i}$ were taken to be stochastic and modeled by
a gaussian process. The latter assumption was not based on analysis of the
underlying string theory. The gaussian process represents a large universality
class for stochastic processes and in this sense is an understandable
assumption.Our purpose is to determine the justification of using gaussian (or
other) distributions for the velocity recoil in the context of D-particle foam.

\bigskip

\section{Moduli and vertex operators for D-particles: a Comprehensive Review \label{sec:3}}

\bigskip

In order to understand the dynamics of D-particle foam it is imperative to
consider D-particle quantisation and in particular D-particle recoil as a
result of scattering off stringy matter. The issue of recoil is not fully
understood and it is not our purpose here to delve into the subtleties of
recoil and operators describing recoil. We will rather proceed on the basis of
a proposal which has had some success in the past \cite{kogan}. The energy of
a D-particle is independent of its position. Consequently in bosonic string
theory there are 25 zero modes, the Dirichlet directions. Zero modes lead to
infrared divergences in loops in a field theory setting where collective
co-ordinates are used to isolate these infrared divergences. This is
essentially due to the naivety of the perturbation series that is used and can
be addressed using a coherent state formalism. The situation is similar but,
in some ways, worse for string theory since the second quantised formalism for
strings is more rudimentary. In string theory the use of first quantisation
requires a sum over Riemann sheets with different moduli parameters. The
formal transition from one surface to another of lower genus (within string
perturbation theory ) has singularities associated with infrared divergences
in the integration over moduli parameters. In the case of a disc $D$ an
incipient annulus $\Sigma$ can be found by making two punctures and attaching
a long thin strip (somewhat like the strap in a handbag and dubbed wormholes).
In the limit of vanishing width this wormhole, is represented by a region in
moduli space, integration over which leads to an infrared divergence. A
similar argument would apply to a Riemann sphere where two punctures would be
connected by a long thin tube. For the disc the divergence will be due to
propagation of zero mode open string state while for the Riemann sphere it
would be an analogous closed string state along the wormhole. Let us examine
this divergence explicitly. Consider the correlation function $\left\langle
V_{1}V_{2}V_{3}\ldots V_{n}\right\rangle _{\Sigma}$ for vertex operators
\cite{green} $\left\{  V_{i}\right\}  _{i=1,\ldots,n}$ where $\Sigma$ is the
Riemann surface of an annulus; it can be deformed into a disc with a wormhole
attached \cite{coll},\cite{szabo}. The set of vertex operators include
necessarily any higher dimensional D-branes necessary to conserve string
charge. However such vertex insertions clearly do not affect the infrared
divergence caused by \ wormwholes. The correlation function can be expressed
as
\begin{equation}
\left\langle V_{1}V_{2}V_{3}\ldots V_{n}\right\rangle _{\Sigma}=\sum_{a}%
\int{\mathrm{{d}}s_{1}}\int{\mathrm{{d}}s_{2}}\int\frac{\mathrm{d}q}%
{q}q^{h_{a}-1}\left\langle \phi_{a}\left(  s_{1}\right)  \phi_{a}\left(
s_{2}\right)  V_{1}V_{2}V_{3}\ldots V_{n}\right\rangle _{D} \label{infrared}%
\end{equation}
where $s_{1}$ and $s_{2}$ are the positions of the punctures on $\partial
D$.The $\left\{  \phi_{a}\right\}  $ are a complete set of eigenstates of the
Virasoro operator $L_{0}$ with conformal weights $h_{a}$\cite{mathieu} and $q$
is a Teichmuller parameter associated with the added thin strip. Clearly there
is a potential divergence associated with its disappearance $q\rightarrow0$
and this corresponds to a long thin strip attached to the disc. For a static
D-particle the string co-ordinates $\overrightarrow{X}_{D}=\left\{
X^{1},X^{2},X^{3},\ldots,X^{25}\right\}  $ have Dirichlet boundary conditions
while $X^{0}$ has Neumann boundary conditions (in the static gauge)%
\begin{equation}
\left.  \frac{\partial}{\partial\sigma}X^{0}\left(  \tau,\sigma\right)
\right\vert _{\sigma=0}=0=\left.  \frac{\partial}{\partial\sigma}X^{0}\left(
\tau,\sigma\right)  \right\vert _{\sigma=\pi}%
\end{equation}
where $\left(  \tau,\sigma\right)  $ is a co-ordinisation of the
worldsheet.The associated translational zero mode is given by%
\begin{equation}
\phi^{i}\left(  X,\omega\right)  =\frac{\sqrt{g_{s}}}{4}\partial_{n}X_{D}%
^{i}\e^{i\omega X^{0}}, \label{zeromode}%
\end{equation}
where $\partial_{n}$ denotes a derivative in the $X^{i}$ Dirichlet direction,
and is an element of the set $\left\{  \phi_{a}\right\}  $. The conformal
weight $h_{i}=1+\alpha^{\prime}\omega^{2}$. The relevant part of the integral
in (\ref{infrared}) is
\begin{equation}
\int_{0}^{1}dq\int_{-\infty}^{\infty}d\omega\,q^{-1+\alpha^{\prime}\omega^{2}%
}=\int_{0}^{1}dq\,\frac{1}{q\left(  -\log q\right)  ^{1/2}} \label{relevant}%
\end{equation}
and is divergent because of the behaviour of the integrand near $q=0.$ This can
be regularized by putting a lower cut-off $q > \delta \to 0$ in the integral.The
correlation function on $\Sigma$ can be computed to be \cite{coll}
\begin{equation}
\left\langle V_{1}V_{2}V_{3}\ldots V_{n}\right\rangle _{\Sigma}=-\frac{g_{s}%
}{16T}\log\delta \left\langle \partial_{n}\overrightarrow{X}_{D}\left(
s_{1}\right)  .\partial_{n}\overrightarrow{X}_{D}\left(  s_{2}\right)
V_{1}V_{2}V_{3}\ldots V_{n}\right\rangle _{D} \label{infred}%
\end{equation}
where $T$ is a cut-off for large (target) time. Division by it removes
divergencies due to the integration over the world-sheet zero modes of the target time. These should not  be confused with divergencies associated with pinched world-sheet surfaces, proportional to ${\rm log}\delta$, that we are interested in here.
These latter divergences cause
conventional conformal invariance to fail.

However, in one approach, it was
argued sometime ago, that these divergences can be canceled if D-particles are
allowed to recoil~\cite{kogan,szabo}, as a result of momentum conservation during their scattering with string states.
In fact, in our D-particle foam model~\cite{Dfoam}, this is not a simple scattering process, as
it involves capture and re-emission of the string state by the D-particle defect.
In simple terms, this process involves splitting of strings by the defect.
In a world-sheet (first quantization) framework, such processes are described by appropriate vertex operators, whose operator product expansion close on a (local) logarithmic algebra~\cite{kogan,szabo}. The
translational zero modes are associated with infinitesimal translations in the
Dirichlet directions. Given that D-particles do not have any internal or
rotational degrees of freedom, these modes should give us valuable information
concerning recoil. Moreover (for $\omega=0$) there is a degeneracy in the
conformal weights between $\phi^{i}$, $\partial_{\omega}\phi^{i}$ and the
identity operator and also the conformal blocks in the corresponding algebra
have logarithmic terms.

To understand the formal structure of the world-sheet deformation operators
pertinent to the recoil/capture process, we first notice that
the world-sheet boundary operator $\mathcal{V}_{\text{D}}$ describing the
excitations of a moving heavy D0-brane is given in the tree approximation by:
\begin{equation}
\mathcal{V}_{\text{D}}=\int_{\partial D}\left(  y_{i}\partial_{n}X^{i}%
+u_{i}X^{0}\partial_{n}X^{i}\right)  \equiv\int_{\partial D}Y_{i}\left(
X^{0}\right)  \partial_{n}X^{i} \label{recoilop}%
\end{equation}
where $u_{i}$ and $y_{i}$ are the velocity \ and position of the D-particle
respectively and $Y_{i}\left(  X^{0}\right)  \equiv y_{i}+u_{i}X^{0}$. To
describe the capture/recoil we need an operator which has non-zero matrix elements between
different states of \ the D-particle and is turned on ``abruptly'' in target time. One way of doing this is to put~\cite{kogan} a $\Theta\left(  X^{0}\right)  $, the Heavyside function, in front of
$\mathcal{V}_{\text{D}}$ which models an impulse whereby the D-particle starts
moving at $X^{0}=0$. Using Gauss's theorem this impulsive $\mathcal{V}%
_{\text{D}}$, denoted by $\mathcal{V}_{\text{D}}^{imp}$, can be represented
as
\begin{equation}
\mathcal{V}_{\text{D}}^{imp}=\sum_{i=1}^{25}\int_{D}d^{2}z\,\partial_{\alpha
}\left(  \left[  u_{i}X^{0}\right]  \Theta\left(  X^{0}\right)  \partial
^{\alpha}X^{i}\right)  =\sum_{i=1}^{25}\int_{\partial D}d\tau\,u_{i}%
X^{0}\Theta\left(  X^{0}\right)  \partial_{n}X^{i}. \label{fullrec}%
\end{equation}
Since $X^{0}$ is an operator it will be necessary to define $\Theta\left(
X^{0}\right)  $ as an operator using the contour integral%
\begin{equation}
\Theta_{\varepsilon}\left(  X^{0}\right)  =-\frac{i}{2\pi}\int_{-\infty
}^{\infty}\frac{d\omega}{\omega-i\varepsilon}\e^{i\omega X^{0}}\text{ with }\varepsilon
\rightarrow0+.
\end{equation}
Hence we can consider%
\begin{equation}\label{Depsilonop}
D_{\varepsilon}(X^0) \equiv D (X^0 ; \varepsilon) = X^{0}\Theta_{\varepsilon}\left(  X^{0}\right)
=-\int_{-\infty}^{\infty}\frac{d\omega}{\left(  \omega -i\varepsilon\right)  ^{2}%
}\e^{i\omega X^{0}}~.
\end{equation}
The introduction of the feature of impulse in the operator breaks conventional
conformal symmetry, but a modified logarithmic conformal algebra holds. A
generic logarithmic algebra in terms of operators $\mathcal{C}$ and
$\mathcal{D}$ and the stress tensor $T\left(  z\right)  $\ (in complex tensor
notation ) satisfies the operator product expansion%
\begin{align}
T\left(  z\right)  \mathcal{C}\left(  w,\overline{w}\right)   &  \sim
\frac{\Delta}{\left(  z-w\right)  ^{2}}\mathcal{C}\left(  w,\overline
{w}\right)  +\frac{\partial\mathcal{C}\left(  w,\overline{w}\right)  }{\left(
z-w\right)  }+\cdots\\
T\left(  z\right)  \mathcal{D}\left(  w,\overline{w}\right)   &  \sim
\frac{\Delta}{\left(  z-w\right)  ^{2}}\mathcal{D}\left(  w,\overline
{w}\right)  +\frac{1}{\left(  z-w\right)  ^{2}}\mathcal{C}\left(  w\right)
+\frac{\partial\mathcal{D}\left(  w\right)  }{\left(  z-w\right)  }+\cdots
\end{align}
and
\begin{align}
\left\langle \mathcal{C}\left(  z,\overline{z}\right)  \mathcal{C}\left(
0,0\right)  \right\rangle  &  \sim0\label{can1}\\
\left\langle \mathcal{C}\left(  z,\overline{z}\right)  \mathcal{D}\left(
0,0\right)  \right\rangle  &  \sim\frac{c}{\left\vert z\right\vert ^{2\Delta}%
}\label{can2}\\
\left\langle \mathcal{D}\left(  z,\overline{z}\right)  \mathcal{D}\left(
0,0\right)  \right\rangle  &  \sim\frac{c}{\left\vert z\right\vert ^{2\Delta}%
}\left(  \log\left\vert z\right\vert +\mathfrak{d}\right)  \label{can3}%
\end{align}
where $\mathfrak{d}$ is a constant. Since the conformal dimension of
$\e^{iqX^{0}}$ is $\frac{q^{2}}{2}$ we find that
\begin{equation}
T\left(  w\right)  D_{\varepsilon}\left(  z\right)  \sim-\frac{\varepsilon
^{2}}{2\left(  w-z\right)  ^{2}}D_{\varepsilon}\left(  z\right)  +\frac
{1}{\left(  w-z\right)  ^{2}}\varepsilon\Theta_{\varepsilon}\left(
X^{0}\right)  +\cdots
\end{equation}
and so a logarithmic conformal algebra structure arises if we define
\begin{equation}\label{Cepsilonop}
C_{\varepsilon} (X^0) \equiv C(X^0 ; \varepsilon)  =\varepsilon\Theta_{\varepsilon}\left(
X^{0}\right)~,
\end{equation}
suppressing, for simplicity, the non-holomorphic piece. The
above logarithmic conformal field theory structure is found with this
identification. Similarly we find%
\[
T\left(  w\right)  C_{\varepsilon}\left(  z\right)  \sim-\frac{\varepsilon
^{2}}{2\left(  w-z\right)  ^{2}}C_{\varepsilon}\left(  z\right)  +\cdots
\]
Consequently $\Delta$ for $C_{\varepsilon}\left(  z\right)  $ and
$D_{\varepsilon}\left(  z\right)  $ is $-\frac{\varepsilon^{2}}{2}$. A
calculation (in a euclidean metric) for a disc of size $L$ with a
short-distance worldsheet cut-off $a$ reveals that as $\varepsilon\rightarrow0$%
\begin{align}
\left\langle C_{\varepsilon}\left(  z\right)  C_{\varepsilon}\left(  0\right)
\right\rangle  &  \sim O\left(  \varepsilon^{2}\right) \label{canep1}\\
\left\langle C_{\varepsilon}\left(  z,\overline{z}\right)  D_{\varepsilon
}\left(  0\right)  \right\rangle  &  \sim\frac{\pi}{2}\sqrt{\frac{\pi
}{\varepsilon^{2}\alpha}}\left(  1-2\varepsilon^{2}\log\left\vert \frac{z}%
{a}\right\vert ^{2}\right) \label{canep2}\\
\left\langle D_{\varepsilon}\left(  z,\overline{z}\right)  D_{\varepsilon
}\left(  0\right)  \right\rangle  &  \sim\frac{\pi}{2}\sqrt{\frac{\pi
}{\varepsilon^{2}\alpha}}\left(  \frac{1}{\varepsilon^{2}}-2\log\left\vert
\frac{z}{a}\right\vert ^{2}\right)  \label{canep3}%
\end{align}
where $\alpha=\log\left\vert \frac{L}{a}\right\vert ^{2}$. We consider
$\varepsilon\rightarrow0+$ such that
\begin{equation}
\varepsilon^{2}\alpha\sim\frac{1}{2\eta}=O\left(  1\right)  ~,
\label{epscutoff}%
\end{equation}
where $\eta$ is the time signature and the right-hand side is kept fixed as
the cutoff runs; it is then straightforward to see that (\ref{canep1}),
(\ref{canep2}), and (\ref{canep3}) are consistent with (\ref{can1}),
(\ref{can2}), and (\ref{can3}). It is only under the condition
(\ref{epscutoff}) that the recoil operators $C_{\varepsilon}$ and
$D_{\varepsilon}$ obey a closed logarithmic conformal algebra~\cite{kogan}:
\begin{align}
<C_{\varepsilon}(z)C_{\varepsilon}(0)>  &  \sim0\nonumber\\
<C_{\varepsilon}(z)D_{\varepsilon}(0)>  &  \sim1\nonumber\\
<D_{\varepsilon}(z)D_{\varepsilon}(0)>  &  \sim-2\eta\log|z/L|^{2} \label{CD}%
\end{align}
The reader should notice that the full recoil operators, involving
$\partial_{n}X^{i}$ holomorphic pieces with the conformal-dimension-one
entering (\ref{fullrec})), obey the full logarithmic algebra (\ref{can1}),
(\ref{can2}), (\ref{can3}) with conformal dimensions $\Delta=1-\frac
{\varepsilon^{2}}{2}$. From now on we shall adopt the euclidean signature
$\eta=1$.

We next remark that, at tree level in the string perturbation sense, the
stringy sigma model (inclusive of the D-particle \ boundary term and other
\ vertex operators) is a two dimensional renormalizable quantum field theory;
hence for generic couplings $g^{i}$ it is possible to see how the couplings
run in the renormalization group sense with changes in the short distance
cut-off through the beta functions $\beta^{i}$. In the world-sheet
renormalization group \cite{klebanov}, based on expansions in powers of the
couplings, $\beta^{i}$ has the form ( with no summation over the repeated
indices)
\begin{equation}
\beta^{i}=y_{i}g^{i}+\ldots\label{beta fn}%
\end{equation}
where $y_{i}$ is the anomalous dimension, which is related to the conformal
dimension $\Delta_{i}$ by $y_{i} = \Delta_{i} - \delta$, with $\delta$ the
engineering dimension (for the holomorphic parts of vertex operators for the
open string $\delta= 1$). The $\ldots$ in (\ref{beta fn}) denote higher orders
in $g^{i}$. Consequently, in our case, we note that the (renormalised)
D-particle recoil velocities $u^{i}$ constitute such $\sigma$-model couplings,
and to lowest order in the renormalised coupling $u_{i}$ the corresponding
$\beta$ function satisfies%
\begin{equation}
\frac{du^{i}}{d\log\Lambda}=-\frac{\varepsilon^{2}}{2}u^{i}. \label{rengp}%
\end{equation}
where $\Lambda$ is a (covariant) world-sheet renormalization-group scale.
In our notation, we
identify the logarithm of this scale with $\alpha=\log\left\vert \frac{L}%
{a}\right\vert ^{2}$, satisfying (\ref{epscutoff}).

An important comment is now in order concerning the interpretation of the flow
of this world-sheet renormalization group scale as a target-time flow. The
target time $t$ is identified through $t=2\log\Lambda$. For completeness we
recapitulate the arguments of \cite{kogan} leading to such a conclusion. Let
one make a scale transformation on the size of the world-sheet
\begin{equation}
L\rightarrow L^{\prime}=\e^{t/4}L \label{fsscaling}%
\end{equation}
which is a finite-size scaling (the only one which has physical sense for the
open string world-sheet). Because of the relation between $\varepsilon$ and
$L$ (\ref{epscutoff}) this transformation will induce a change in
$\varepsilon$
\begin{equation}
\varepsilon^{2}\rightarrow\varepsilon^{\prime2}=\frac{\varepsilon^{2}%
}{1+\varepsilon^{2}t} \label{epsilontransform}%
\end{equation}
(note that if $\varepsilon$ is infinitesimally small, so is $\varepsilon^{\prime}$
for any finite $t$). From the scale dependence of the correlation functions
(\ref{CD}) that $C_{\varepsilon}$ and $D_{\varepsilon}$ transform as:
\begin{align}
D_{\varepsilon}  &  \rightarrow D_{\varepsilon^{\prime}}=D_{\varepsilon
}+tC_{\varepsilon}\nonumber\\
C_{\varepsilon}  &  \rightarrow C_{\varepsilon^{\prime}}=C_{\varepsilon}%
\end{align}
From this transformation one can then see that the coupling constants in front
of $C_{\varepsilon}$ and $D_{\varepsilon}$ in the recoil operator
(\ref{recoilop}), i.e. the velocities $u_{i}$ and spatial collective
coordinates $y_{i}$ of the brane, must transform like:
\begin{equation}
u_{i}\rightarrow u_{i}~~,~~y_{i}\rightarrow y_{i}+u_{i}t \label{scale2}%
\end{equation}
This transformation is nothing other but the Galilean
transformation for the heavy D-particles and thus it demonstrates that the finite size scaling parameter $t$, entering
(\ref{fsscaling}), plays the r\^ole of target time, on account of (\ref{epscutoff}). Notice that (\ref{scale2}) is derived upon using (\ref{CD}), that is in the limit where $\varepsilon \to 0$.
This will become important later on, where we shall discuss (stochastic) relaxation phenomena in our recoiling D-particle.

Thus, in the presence of recoil a world-sheet scale transformation leads to an
evolution of the $D$-brane in target space, and from now on we identify the
world-sheet renormalization group scale with the target time $t$. In this
sense, equation (\ref{rengp}) is an evolution equation in target time.

However, this equation does not capture quantum-fluctuation aspects of $u^{i}$ about its classical trajectory with time $u_i(t)$. Going
to higher orders in perturbation theory of the quantum field theory at fixed
genus does not qualitatively alter the situation in the sense that the
equation remains deterministic. In the next section we shall consider the
effect of string perturbation theory where higher genus surfaces are
considered and re-summed in some appropriate limits that we shall discuss in detail.

\bigskip

\section{String perturbation theory and implication for recoil velocity \label{sec:4}}

\bigskip

It is not possible to exactly sum up higher orders in string perturbation
theory. We have seen that infrared singularities in the integration over the
moduli of the Riemann surface (representing the world sheet) in the wormhole
limit are related to the recoil operators for the D-particle. The wormhole
construction \cite{coleman} is a way of constructing higher genus surfaces
from lower genus ones. Since it will be relevant to us later, we should note
that $g_{s}$ the string coupling is given by
\begin{equation}
g_{s}=\mathrm{e}^{\left\langle \Phi\right\rangle } \label{string coupling}%
\end{equation}
where $\Phi$ is the spin zero dilaton mode which is part of \ the massless
string multiplet. Here $\left\langle \ldots\right\rangle $ denotes the string
path integral $\int DX\,\mathrm{e}^{S_{\sigma}}\Phi$ where $S_{\sigma}$ is the
string $\sigma$-model action in the presence of string backgrounds such as the
dilaton and the Kalb-Ramond modes. In particular the $\sigma$ model
deformation due to the dilaton has the form
\begin{equation}
\frac{1}{4\pi}\int_{\Sigma}d\sigma d\tau\,\sqrt{\gamma}\,\Phi\left(  X\right)
R^{\left(  2\right)  }\left(  \tau,\sigma\right)  \label{dilaton}%
\end{equation}
on a worldsheet Riemann surface $\Sigma$ where $\gamma_{\alpha\beta}$ is the
induced metric on the worldsheet, $\gamma=\left\vert \det\gamma_{\alpha\beta
}\right\vert $ and \ $R^{\left(  2\right)  }$ is the associated Ricci
curvature scalar. Now the Euler characterisitic $\chi$ of $\Sigma$ is given by%
\begin{equation}
\chi=\frac{1}{4\pi}\int_{\Sigma}d\sigma d\tau\sqrt{\gamma}R^{\left(  2\right)
}=2\left(  1-g\right)  \label{euler}%
\end{equation}
where $g$ is the genus and is an integer valued invariant. If we split the
dilaton into a classical (worldsheet co-ordinate independent) part
$\left\langle \Phi\right\rangle $ and a quantum part $\varphi=\colon\Phi
\colon$, where $\colon\ldots\colon$ denotes appropriate normal ordering, we
can write $\Phi=\left\langle \Phi\right\rangle +\varphi$. The $\sigma$-model
partition function $Z$\ can be written as a sum over genera
\begin{align}
Z  &  =\sum_{\chi}\int\int d\gamma_{\alpha\beta}\,dX\,\mathrm{e}%
^{-S_{rest}-\chi\left\langle \Phi\right\rangle -\frac{1}{4\pi}\int_{\Sigma
}d\sigma d\tau\,\sqrt{\gamma}\,\varphi R^{\left(  2\right)  }\left(
\tau,\sigma\right)  }\label{partiition1}\\
&  =\sum_{\chi}g_{s}^{-\chi}\int\int d\gamma_{\alpha\beta}\,dX\,\mathrm{e}%
^{-S_{rest}-\frac{1}{4\pi}\int_{\Sigma}d\sigma d\tau\,\sqrt{\gamma}\,\varphi
R^{\left(  2\right)  }\left(  \tau,\sigma\right)  } \label{partition2}%
\end{align}
where $S_{rest}$ denotes a $\sigma$-model action involving the rest of the
background deformations except the dilaton. For the moment we will assume that
the theory is such that a potential is generated for $\Phi$ which suppresses
the fluctuations represented by $\varphi$. In general we would have to
consider $g_{s}=\mathrm{e}^{\Phi}$ which would then make the string coupling a field.

The summation over genera cannot be performed exactly. We will follow an
approach using a mechanism due to Fischler and Susskind \cite{fs},\cite{szabo}
based on a dilute gas of wormholes (proposed originally by Coleman within the
context of euclidean quantum gravity \cite{coleman}). This results in the
structure of recoil (in lowest order) being modified by generating a gaussian
distribution for the recoil velocity $u^{i}$.

We present a detailed review of the pertinent formalism in Appendix A. For our
purposes in this section we note that, in the case of mixed logarithmic
states, the pinched topologies are characterized by divergences of a double
logarithmic type (c.f. (\ref{CDprop}) in Appendix A) which arise from the form
of the string propagator (c.f. (\ref{bilocal}) in Appendix A) in the presence
of generic logarithmic operators $C$ and $D$, $\int dq~q^{\Delta_{\varepsilon
}-1}\,\langle C,D|%
\begin{pmatrix}
1 & \log q\cr0 & 1\cr
\end{pmatrix}
|C,D\rangle$~.

As shown in \cite{kogan}, the mixing between $C$ and $D$ states along
degenerate handles leads formally to divergent string propagators in physical
amplitudes, whose integrations have leading divergences of the form $\int
\frac{dq}q~\log q\int d^{2}z~D(z;\varepsilon)\int d^{2}z^{\prime}~C(z^{\prime2
}\int d^{2}z~D(z;\varepsilon)\int d^{2}z^{\prime}~C(z^{\prime};\varepsilon
)$~.As explained in \cite{szabo}, and reviewed in Appendix A of the current
manuscript, these $(\log\delta)^{2}$ divergences can be cancelled by imposing
momentum conservation in the scattering process of the light string states off
the D-particle background.

We note at this stage that, as mentioned earlier, isolated D-particles do not
exist, as a result of their gauge flux conservation requirement. The
physically correct way to formulate, therefore, the problem, is to consider
groups of $N$, say, D-particles, which interact among themselves with
flux-carrying stretched strings. The above analysis remains intact (in the
sense of generalising straightforwardly) when more than one D-particle is
present in a region with typical dimensions smaller than the string length
sometimes known as the fat brane \cite{szabo}. There are, of course, technical
differences in the sense that a non-abelian structure arises and the
$\widehat{Y}$ have matrix labels. All qualitative features, however, are
preserved concerning the quantisation of the D-particle background moduli. For
the remainder of this section we shall, therefore, formulate our arguments
within this rigorous multi-D-particle picture.

The cancelation of leading divergences of the genus expansion in the
non-abelian case of a group of $N$ D-particles, is demonstrated explicitly in
appendix A. It is shown there that this renormalization requires that the
change in (renormalized) velocity of the due to the recoil from the scattering
of string states be
\begin{align}
\bar U_{i}^{ab}=-\frac1{M_{s}}\,\Bigl(k_{1}+k_{2}\Bigr)_{i}\,\delta^{ab}%
=\frac{d\bar Y_{i}^{ab}}{dt}~, \qquad a,b = 1 \dots N \label{recoilvel}%
\end{align}
where $k_{1,2}$ are the initial and final momenta in the scattering process
and $M_{D}=1/\sqrt{\alpha^{\prime}}\, g_{s}$ is the BPS mass of the string
soliton \cite{polch,polch2}, and $g_{s} < 1 $ is the physical (weak) string
coupling. In (\ref{recoilvel}), the $k_{1,2}$ are true physical momenta so
that $M_{D}$ represents the actual BPS mass of the D-particles. This means
that, to leading order, the constituent D-particles in a group of $N$ of them,
say, move parallel to one another with a common velocity and there are no
interactions among them. Thus the leading recoil effects imply a commutative
structure and the ``fat brane'' of the group of D-particles behaves as a
single D-particle (with a single average collective coordinate of its center
of mass). In such a limit one may replace $U_{i}^{ab}$ by $u_{i}$ (c.f.
previous section), describing the collective recoil velocity of the fat brane.
This should be understood throughout this work.

In addition to this divergence, there are sub-leading $\log\delta$
singularities, corresponding to the diagonal terms $\int d^{2}%
z~D(z;\varepsilon)\int d^{2}z^{\prime}~D(z^{\prime};\varepsilon)$ and $\int
d^{2}z~C(z;\varepsilon)\int d^{2}z^{\prime}~C(z^{\prime};\varepsilon)$. These
latter terms are the ones we should concentrate upon for the purposes of
deriving the quantum fluctuations of the collective D-particle coordinates. It
is these sub-leading divergences in the genus expansion which lead to
interactions between the constituent D-branes and provide the appropriate
noncommutative quantum extension of the leading dynamics (\ref{recoilvel}).
The reader should recall that these (sub-leading) divergences also showed up
in the much simpler case of perpetual Galilean motion of D-branes discussed in
\cite{coll} (c.f. (\ref{infred})), as a result of the translational symmetries
zero mode contributions.

In the weak-coupling case, we can truncate the genus expansion to a sum over
pinched annuli (fig. \ref{fig:annulus} in Appendix A). This truncation
corresponds to a semi-classical approximation to the full quantum string
theory in which we treat the D-particles as heavy non-relativistic objects in
target space. Then the dominant contributions to the sum are given by the
$\log\delta$ modular divergences described above, and the effects of the
dilute gas of wormholes on the disc are to \emph{exponentiate} the bilocal
operator (\ref{bilocal}) of Appendix A, describing string propagation in a
pinched annulus. Thus, in the pinched approximation, the genus expansion of
the bosonic $\sigma$-model leads to an effective change in the matrix $\sigma
$-model action by~\cite{szabo}
\begin{align}
\Delta S\simeq\frac{g_{s}^{2}}2\log\delta\sum_{a,b,c,d}\,\int_{-\infty
}^{\infty}d\omega~d\omega^{\prime}~\oint_{\partial\Sigma}\oint_{\partial
\Sigma^{\prime}} V_{ab}^{i}(x;\omega)~G_{ij}^{ab;cd}(\omega,\omega^{\prime
})~V_{cd}^{j}(x;\omega^{\prime}) \label{actionchange}%
\end{align}
where $\omega, \omega^{\prime}$ are Fourier variables, defined appropriately
in Appendix A, and $G_{ij}$~, $i,j = C,D$ is a metric in the theory space of
strings, introduced by Zamolodchikov~\cite{zam}.

The bilocal action (\ref{actionchange}) can be cast into the form of a local
worldsheet effective action by using standard tricks of wormhole calculus
\cite{coleman} and rewriting it as a functional Gaussian
integral~\cite{szabo}
\begin{align}
\mathrm{e}^{\Delta S}  &  =\int[d\breve{\rho}]~\exp\left[  -\frac
12\sum_{a,b,c,d}\,\int_{-\infty}^{\infty}d\omega~d\omega^{\prime}~\breve{\rho
}_{i}^{ab}(\omega)~\oint_{\partial\Sigma} \oint_{\partial\Sigma^{\prime}%
}G^{ij}_{ab;cd}(\omega,\omega^{\prime})~ \breve{\rho}_{j}^{cd}(\omega^{\prime
})\right. \nonumber\\
&  \left.  ~~~~~~~~~~~~~~~~~~~~ +\,g_{s}\,\sqrt{\log\delta}~\sum_{a,b=1}%
^{N}\,\int_{-\infty}^{\infty}d\omega~\breve{\rho}_{i}^{ab}(\omega
)\,\oint_{\partial\Sigma} V_{ab}^{i}(x;\omega)\right]  \label{Gaussianint}%
\end{align}
where $\breve{\rho}_{i}^{ab}(\omega)$ are stochastic coupling constants of the
worldsheet matrix $\sigma$-model, which express quantum fluctuations of the
corresponding background fields in target space, as a consequence of genus
re-summation. Thus the effect of the resummation over pinched genera is to
induce quantum fluctuations of the collective D-brane background, leading to a
set of effective quantum coordinates
\begin{align}
\breve{Y}_{i}^{ab}(\omega)~\to~\widehat{\mathcal{Y}}_{i}^{ab}(\omega
)=\breve{Y}_{i}^{ab}(\omega)+g_{s}\,\sqrt{\log\delta}~ \breve{\rho}_{i}%
^{ab}(\omega) \label{qcoupling}%
\end{align}
viewed as position operators in a co-moving target space frame.

Thus we find that the genus expansion in the pinched approximation
is~\cite{szabo}
\begin{align}
\sum_{h^{(p)}}Z_{N}^{h^{(p)}}[A]~\simeq~\left\langle \int_{\mathcal{M}}%
[d\rho]~\wp[\rho] {}~W\!\left[  \partial\Sigma
;A-\mbox{$\frac1{2\pi\alpha'}$}\,\rho\right]  \right\rangle _{0}
\label{pinchedpartfn}%
\end{align}
where the sum is over all pinched genera of infinitesimal pinching size, and
\begin{align}
\wp[\rho] \propto\exp\left[  -\frac1{2\Gamma^{2}}\sum_{a,b,c,d}\,\int_{0}%
^{1}ds~ds^{\prime}~ \rho_{i}^{ab}\left(  X^{0}(s)\right)  \,G_{ab;cd}%
^{ij}(s,s^{\prime})\,\rho_{j}^{cd} \left(  X^{0}(s^{\prime})\right)  \right]
\label{Gaussiandistr}%
\end{align}
is a (appropriately normalized) functional Gaussian distribution on moduli
space of width
\begin{align}
\Gamma=g_{s}\,\sqrt{\log\delta} \label{widthdef}%
\end{align}
In (\ref{pinchedpartfn}) we have normalized the functional Haar integration
measure $[d\rho]$ appropriately.

We see therefore that the diagonal sub-leading logarithmic divergences in the
modular cutoff scale $\delta$, associated with degenerate strips in the genus
expansion of the matrix $\sigma$-model, can be treated by absorbing these
scaling violations into the width $\Gamma$ of the probablity distribution
characterizing the quantum fluctuations of the (classical) D-brane
configurations $Y_{i}^{ab}(X^{0}(s))$. In this way the interpolation among
families of D-brane field theories corresponds to a quantization of the
worldsheet renormalization group flows. Note that the worldsheet wormhole
parameters, being functions on the moduli space of recoil deformations, can be
decomposed as
\begin{align}
\rho_{i}^{ab}(X^{0}(s))=\lim_{\varepsilon\to0^{+}}\left(  [\rho_{C}]_{i}%
^{ab}C(X^{0};\varepsilon) +[\rho_{D}]_{i}^{ab}D(X^{0};\varepsilon)\right)
\label{wormholedecomp}%
\end{align}
The fields $\rho_{C,D}$ are then renormalized in the same way as the D-brane
couplings, so that the corresponding renormalized wormhole parameters generate
the same type of (Galilean) $\beta$-function equations (\ref{rengp}).

According to the standard Fischler-Susskind mechanism for canceling string
loop divergences \cite{fs}, modular infinities should be identified with
worldsheet divergences at lower genera. Thus the strip divergence $\log\delta$
should be associated with a worldsheet ultraviolet cutoff scale $\log\lambda$,
which in turn is identified with the target time as described earlier.

We may in effect take $\delta$ independent from $\Lambda$, in which case we
can first let $\varepsilon\to0^{+}$ in the above and then take the limit
$\delta\to0$. Interpreting $\log\delta$ in this way as a renormalization group
time parameter (interpolating among D-brane field theories), the time
dependence of the renormalized width (\ref{widthdef}) expresses the usual
properties of the distribution function describing the time evolution of a
wavepacket in moduli space. The inducing of a statistical Gaussian spread of
the D-brane couplings is the essence of the quantization procedure.

A final remark is in order. From the form (\ref{Depsilonop}) and
(\ref{Cepsilonop}) of the recoil operators, it is evident that the dominant
contributions in the limit $\varepsilon\to0^{+}$, we consider here, come from
the $D$-deformations, pertaining to the recoil velocity $u^{i}$ of the
D-particle (or, better, the center of mass velocity of a group of D-particles,
as discussed above). From now on, therefore, we restrict our attention to the
distribution functions of such recoil velocities:
\begin{equation}
\wp(u) \sim\frac{1}{\Gamma} \mathrm{e}^{-\frac{u^{2} - {\bar u}^{2}}%
{\Gamma^{2}}}~, \qquad\Gamma= g_{s} \sqrt{\mathrm{log}\delta}~,
\label{gaussian}%
\end{equation}
where $\bar u$ denotes the classical recoil velocity. Notice that, upon
invoking~\cite{szabo} the Fischler-Susskind mechanism~\cite{fs} for the
absorption of the modular infinities to lower-genus (disc) world-sheet
surfaces, we may identify $\mathrm{log}\delta$ with the target time:
\begin{equation}
\mathrm{log}\delta= t~, \label{modtime}%
\end{equation}
where this identification should be understood as being implemented at the end
of the computation. To be precise, as explained in \cite{szabo}, the correct
form of (\ref{modtime}) would be: $\mathrm{log}\delta= g_{s}^{\chi}~ t$, with
$\chi> 0$ an exponent that can only be determined phenomenologically in the
approach of \cite{szabo}, by comparing the space-time uncertainty principles,
derived in this approach of re-summing world-sheet genera, with the ones
within standard string/brane theory. In fact, in our approach of re-summing
world-sheet pinched surfaces~\cite{szabo}, one obtains for the spatial and
temporal variances: $\Delta Y^{aa} \Delta t \ge g_{s}^{\chi}\sqrt
{\alpha^{\prime}}$, which implies that the standard string-theory
result~\cite{yoneya}, independent of the string coupling, is obtained for
$\chi= 0$. This is the case we shall consider here, which leads to the
identification (\ref{modtime}). However, in the modern approach of D-brane
theories, one can adjust the uncertainty relations in order to probe minimal
distances below the string length, which is achieved by the
choice~\cite{szabo}, e.g. $\chi= 2/3$, reproducing the characteristic minimal
length probed by D-particles~\cite{liyoneya}. In our case, where, as we shall
discuss in the next subsection, the coupling constant of the string may itself
fluctuate, it is the mean value of $g_{s}$ that enters in such relations. This
issue is not relevant if we stay within the $\chi= 0$ case, which we do in
this article.

We next remark that the nature of the Gaussian correlation is assumed to be
delta correlated in time. The Langevin equation \cite{gardiner2} implied by
(\ref{qcoupling}) replaces (\ref{rengp}) and can be written as
\begin{equation}
\frac{d{\bar{u}}^{i}}{dt}=-\frac{1}{4t}{\bar{u}}^{i}\,+\frac{g_{s}}%
{\sqrt{2\alpha^{\prime}}}t^{1/2}\xi\left(  t\right)  \label{Langevin}%
\end{equation}
where $t=\varepsilon^{-2}$ and $\xi\left(  t\right)  $ represents white noise.
This equation is valid for large $t$. From the above analysis it is known that
\cite{szabo} (c.f. Appendix A)) that to $O\left(  g_{s}^{2}\right)  $ the
correlation for $\xi\left(  t\right)  $ is $\bar{u}^{i}$ independent, \ and
for time scales of interest, is correlated like white noise ; hence the
correlation of $\xi\left(  t\right)  $ has the form:
\begin{equation}
\left\langle \xi\left(  t\right)  \xi\left(  t^{\prime}\right)  \right\rangle
=\delta\left(  t-t^{\prime}\right)  .
\end{equation}
Since the vectorial nature of $\bar{u}^{i}$ is not crucial for our analysis we
will suppress it and consider the single variable $\bar{u}$.

We should stress that this equation is valid for large $t$ which is required
since $\varepsilon$ is small. Hence the apparent singularity in
Eqn(\ref{Langevin} ) at $t=0$ is not relevant and so we can empirically
\emph{regularise} this singularity by changing $\frac{1}{t}$ to $\frac
{1}{t+t_{0}}$ for some $t_{0}>0$; $t_{0}$ is the order of the capture time of
the $\phi$ meson by the D-particle The stochastic Langevin equation
(\ref{Langevin}), describes relaxation aspects of the recoiling D-particle
with equilibrium being reached only as $\varepsilon\rightarrow0$ (or
$t\rightarrow\infty$).

The reader should notice that in the limit the system reaches equilibrium with
a constant in time velocity. It is only in this limit that the Galilean
transformation (\ref{scale2}) applies, as already discussed there. We now
proceed to a solution of this Langevin equation and a discussion on the
pertinent physical consequences for a statistical population of
quantum-fluctuating D-particles.

\bigskip

\section{Solution of Langevin equations and fluctuating string coupling \label{sec:5}}

\bigskip

In the recent modeling of the omega effect~\cite{bernabeu}, the recoil
velocity of the D-particles $u$ has been taken as a classical stochastic
variable. The D-particle fluctuations which are \ described by (\ref{Langevin}%
) will be superimposed on this stochasticity. Eqn.~(\ref{Langevin}) is \ particulary
simple equation in the sense that the drift and diffusion terms are
independent of $u$. By making a change of variable it is easy to eliminate the
drift term and the resulting equation can then be interpreted in terms of a
Wiener process~\cite{gardiner2}. Let us consider the auxiliary equation
\begin{equation}
\frac{d}{dt}y=-\frac{1}{4\left(  t+t_{0}\right)  }y, \label{drift}%
\end{equation}
which just deals with the drift part of Eqn.(\ref{Langevin}). It has a
solution
\[
y\left(  t\right)  =y\left(  t_{0}\right)  \Upsilon\left(  t\right)
\]
where
\begin{equation}
\Upsilon\left(  t\right)  =\exp\left[  -\frac{1}{4}\int_{0}^{t}\frac
{dt^{\prime}}{t^{\prime}+t_{0}}\right]  =\left(  \frac{t+t_{0}}{t_{0}}\right)
^{-\frac{1}{4}}%
\end{equation}
and $t_{0}$ is a time much smaller than $t$. We now define $U\left(  t\right)
=u\left(  t\right)  \Upsilon\left(  t\right)  ^{-1}$ and readily find that
\begin{equation}
\frac{dU}{dt}=\frac{g_{s}}{\sqrt{2\alpha^{\prime}}}t^{1/2}\Upsilon\left(
t\right)  ^{-1}\xi\left(  t\right)  . \label{Langevin2}%
\end{equation}
This describes purely diffusive motion and is thus related to the Wiener
process; equivalently we can consider the associated probability distribution
$p\left(  U,t\right)  $ which satisfies the Fokker-Planck equation%
\begin{equation}
\frac{\partial}{\partial t}p\left(  U,t\right)  =\frac{1}{4\alpha^{\prime}%
}g_{s}^{2}t(\frac{t+t_{0}}{t_{0}})^{\frac{1}{2}}\frac{\partial^{2}}{\partial
U^{2}}p\left(  U,t\right)  . \label{Fokker-Planck}%
\end{equation}
If at $t=0$ consider a D-particle velocity recoil $u_{0}$ so that
\begin{equation}
p\left(  U,0\right)  =\delta\left(  U-u_{0}\right)  ~. \label{initial}%
\end{equation}
The Eqn. (\ref{Fokker-Planck}) can be solved to give
\begin{equation}
p\left(  U,t\right)  =\sqrt{\frac{15\alpha^{\prime}}{2\pi\eta\left(  t\right)
}}\frac{1}{g_{s}}\exp\left(  -\frac{15\alpha^{\prime}(U-u_{0})^{2}}{2g_{s}%
^{2}\eta\left(  t\right)  }\right)  \label{gaussian1}%
\end{equation}
where%
\begin{equation}
\eta\left(  t\right)  =2t_{0}^{2}+3\left(  t+t_{0}\right)  ^{2}\sqrt{1+\frac
{t}{t_{0}}}-5t_{0}^{\frac{1}{2}}(t+t_{0})^{\frac{3}{2}}. \label{dispersion}%
\end{equation}
If the D-particle is typically interacting with matter on time scales of
$t_{0}$, then the effect of a large number of such collisions can be
calculated by performing an ensemble average over a distribution of $u_{0}$. A
distribution for $u_{0}$ that has been used in modelling is a gaussian with
zero mean and variance $\sigma$. This is readily seen to lead to an averaged
distribution D-particle velocity recoil distribution $\ll p\left(  u\left\vert
g_{s}\right.  \right)  \gg$ where
\begin{align}
&  \ll p\left(  u\left\vert g_{s}\right.  \right)  \gg\label{gaussian2}\\
&  =\sqrt{\frac{15\alpha^{\prime}}{2\pi\left(  g_{s}^{2}\eta\left(  t\right)
+15\alpha^{\prime}\sigma^{2}\right)  }}\exp\left[  -\frac{15\alpha^{\prime}%
}{2\left(  g_{s}^{2}\eta\left(  t\right)  +15\alpha^{\prime}\sigma^{2}\right)
}u^{2}\right]  ~.\nonumber
\end{align}
 We have used a notation for $p$ which emphasises that it is conditional on
$g_{s}$ having a fixed value.
The interaction time
includes \emph{both} the time for capture and re-emission of the string by the
D-particle, as well as the time interval until the next capture, during string propagation.
In a generic situation, this time could be much larger than
the capture time, especially in dilute gases of D-particles, which include less than one D-particle per string ($\alpha^{3/2}$) volume. Indeed, as discussed
in detail in \cite{emnuncertnew}, using generic properties of strings
consistent with the space-time uncertainties~\cite{yoneya}, the capture and
re-emission time $t_{0}$, involves the growth of a stretched
string between the string state and the D-brane world (c.f.
fig.~\ref{fig:recoil}) and is found proportional to the incident string energy
$p^{0}$:
\begin{equation}
t_{0} \sim\alpha^{\prime}p^0  \ll \sqrt{\alpha ^{\prime}}~.
\end{equation} We shall use this result in section \ref{sec:6}, where we estimate the strength of the $\omega$-effect
in the initial entangled state of two mesons
(\ref{omega}),  after  the $\phi$-meson decay in
the presence of D-particles. In such a situation, the interaction time is essentially the
capture time $t_0$.

We will now examine how the above results are modified when $g_{s}$
fluctuates. Such issues cannot currently be treated with any level of rigour
since they embody issues of string vacua and backgrounds. Hence we will take a
somewhat phenomenological stance and consider the effect of a class of
stochastic fluctuations for $g_{s}^{-2}\left(  =\mathrm{e}^{-2\left\langle
\Phi\right\rangle }\right)  $ which are varying on a timescale which is slow
compared to the drift time-scale. This is the arena of superstatisitcs
\cite{beck2}. Following the analysis of Beck \cite{beck} (c.f. Appendix B) we
can make the following fairly general classical ansatz for $\frac{1}{g_{s}%
^{2}}$ compatible with its positivity viz.%
\begin{equation}
\frac{1}{g_{s}^{2}}=\sum_{i=1}^{n}x_{i}^{2}\label{ansatz}%
\end{equation}
where the $x_{i}$ are $n$ independent gaussian variables of zero mean and
variance $\sigma_{0}^{2}$. The probability distribution $p$ for $\frac
{1}{g_{s}^{2}}$ is $\chi^{2}$ with $n$ degrees of freedom, i.e.
\begin{equation}
\mathfrak{p}\left(  \frac{1}{g_{s}^{2}}\right)  =\frac{1}{\Gamma\left(
\frac{n}{2}\right)  }\left\{  \frac{ng_{0}^{2}}{2}\right\}  ^{n/2}\left(
\frac{1}{g_{s}^{2}}\right)  ^{\frac{n}{2}-1}\exp\left(  -\frac{ng_{0}^{2}%
}{2g_{s}^{2}}\right)  .\label{chi}%
\end{equation}
Here $\left\langle \frac{1}{g_{s}^{2}}\right\rangle =\frac{1}{g_{0}^{2}%
}$ and the variance $var\left(  \frac{1}{g_{s}^{2}%
}\right)  =\frac{2}{ng_{0}^{4}}$. If $g_{0}$ is held constant the variance can
be made small for large $n$. This is a particular form of the Gamma
distribution. However other choices for $p$, such as lognormal and $F$
distributions have also been considered \cite{beck2} in contexts such as
turbulence. We note that
\begin{equation}
\int_{0}^{\infty}e^{-wu^{2}\beta}\beta^{\frac{n+m}{2}-1}\exp\left(
-\frac{n\beta}{2\beta_{0}}\right)  d\beta=\left(  u^{2}w+\frac{n}{2\beta_{0}%
}\right)  ^{-\frac{m+n}{2}}\Gamma\left(  \frac{m+n}{2}\right)
\end{equation}
We now calculate the probability $p$ is for $u$ as
\begin{equation}
\boldsymbol{p}\left(  u\right)  \equiv\int_{0}^{\infty}d\left(  \frac{1}%
{g_{s}^{2}}\right)  \mathfrak{p}\left(  \frac{1}{g_{s}^{2}}\right)  \ll
p\left(  u\left\vert g_{s}\right.  \right)  \gg.\label{probdefn}%
\end{equation}
The choice of $p$ in (\ref{chi}) leads to canonical form of distribution for
non-extensive statistics \cite{tsallis},\cite{beck} . Other choices mentioned
above give different superstatistics. These wider classes of statistics
present opportunities in interpreting the data in neutrino physics (see e.g.
\cite{mavsarkar}). For large $n$ it is straightforward to show that
\begin{equation}
\boldsymbol{p}\left(  u\right)  =\sqrt{\frac{15\alpha^{\prime}}{\pi\left(
g_{0}^{2}\eta\left(  t\right)  +15\alpha^\prime\sigma^{2}\right)  }}\frac
{\Gamma\left(  \frac{n+1}{2}\right)  \left(  ng_{0}^{2}\right)  ^{n/2}}%
{\Gamma\left(  \frac{n}{2}\right)  \left(  \frac{\boldsymbol{15}
\alpha^{\prime}g_{0}^{2}}{\eta\left(  t\right)  g_{0}^{2}+15\alpha^\prime\sigma^{2}}%
u^{2}+ng_{0}^{2}\right)  ^{\frac{n+1}{2}}}.\label{nonextq}%
\end{equation}
\ On writing $q=1+\frac{2}{n+1}$ and $\widetilde{g}\left(  t\right)
^{2}=\left(  3-q\right)  \left(  \sigma^{2}+\frac{\eta\left(  t\right)
g_{0}^{2}}{15\alpha^\prime}\right)  $ we find the canonical form for Tsallis
statistics\cite{tsallis}
\begin{equation}
\boldsymbol{p}\left(  u\right)  \sim\sqrt{\frac{3-q}{15\widetilde{g}\left(
t\right)  ^{2}}}\left(  \left[  1+\frac{(q-1)u^{2}}{\widetilde{g}\left(
t\right)  ^{2}}\right]  ^{-\frac{1}{q-1}}\right)  .\label{nonextq2}%
\end{equation}
~with $q$ the non-extensivity parameter. Hence the stochasticity in the recoil
velocity, when stringy matter is captured by the D-particle, leads to a
deviation \ from nonextensive Tsallis statistics. The deviation is suppressed
as the interaction time increases. For large $n$ we have weak non-extensivity and the
fluctuations for $\frac{1}{g_{s}^{2}}$ are small.

The quantity of interest which is an important  input for our estimate of the
omega effect is the variance of $\boldsymbol{p}\left(  u\right)  $:
\begin{equation}\label{uvar}
{var}(u)_{\mathrm{non-ext}}=\frac{\eta\left(  t\right)  g_{0}^{2}%
+15\alpha^{\prime}\sigma^{2}}{15\alpha^{\prime}\left(  1-\frac{2}{n}\right)  }.
\end{equation}
For reasons mentioned above, we can make the plausible assumption that the initial state of the neutral K
meson pairs is governed by the variance at $t\sim t_{0}  \sim \alpha ' p^0$.
In such a case, from (\ref{dispersion}) we obtain that $\eta\left(t \sim t_0\right) \simeq 2(1 + \sqrt{2})t_0^2 \sim 2(1 + \sqrt{2})(\alpha^{\prime} p^0)^2 $, and
hence from (\ref{uvar}) the variance over such time scales becomes of order:
\begin{equation}\label{nonext}
{var}(u)_{\mathrm{non-ext}}\left(t\sim t_0 \right) \simeq \frac{2(1 + \sqrt{2}) [g_0^2(\sqrt{\alpha^\prime} p^0)^2 + 15 \sigma^2]}{15(1 - \frac{2}{n})} \sim g_0^2 \alpha^\prime (p^0)^2 (1 + \frac{2}{n} + \dots) + \mathcal{O}(\sigma^2)
\end{equation}
where on the right hand side of the above equation we only gave
an order of magnitude estimate,
assuming that $n$ is large.
The terms of order $\sigma^2$ have not been written explicitly. Indeed, in most models of quantum gravity, a natural assumption would be
that $\sigma^2 \le g_0^2{\alpha^\prime}(p^0)^2 $, which is a natural assumption to make for a dispersion
due to (quantum) fluctuations of the recoil velocity of heavy D-particles of average mass $M_s/g_0 = 1/(g_0\sqrt{\alpha^\prime})$, where $\langle 1/g_s^2 \rangle = 1/g_0^2 $.
As we shall discuss in the next section, therefore, such dispersion terms do not lead to dominant contributions to the $\omega$-effect estimates
which are of primary interest to us here.
These considerations lead to
natural estimates for the parameter $\zeta$ in the magnitude of the omega
effect that we have discussed in our earlier work~\cite{bernabeu}, and we now proceed to examine.

\section{Decoherence and Entangled States: $\omega$-Effect Revisited \label{sec:6}}

We shall use a low-energy quantum-mechanical approach for the dynamics of the
neutral mesons, which is sufficient for a discussion of the $\mathbb{\omega}%
$-effect in non-relativistic systems of entangled mesons, such as Kaons in a
$\phi$-factory~\cite{bernabeu}. For a description of the stochastically
fluctuating space-time effects, we shall make use of the above-derived string
theory effects, in particular the recoil-velocity dispersion (\ref{nonext}),
including the effects of non-extensive statistics (\ref{nonextq2}), due to
fluctuations of the string coupling $g_{s}$.

Following \cite{bernabeu}, where we refer the interested reader for details,
we consider the following interaction Hamiltonian, which expresses the
effective low-energy interaction of the meson states with the D-particle foam
space-time background in the model of \cite{Dfoam}:
\begin{equation}
\widehat{H}=g^{01}\left(  g^{00}\right)  ^{-1}\widehat{k}-\left(
g^{00}\right)  ^{-1}\sqrt{\left(  g^{01}\right)  ^{2}{k}^{2}-g^{00}\left(
g^{11}k^{2}+m^{2}\right)  } \label{GenKG}%
\end{equation}
where $\widehat{k}$ indicates the appropriate momentum operator, with
eigenvalue $k$ (along the direction of motion), when acting on momentum
eigenstates, i.e. $\widehat{k}\left|  \pm k,\uparrow\right\rangle =\pm
k\left|  k,\uparrow\right\rangle $ together with the corresponding relation
for $\downarrow$. The arrows indicate the appropriate meson
``flavours''~\cite{bernabeu}. The induced metric $g_{\mu\nu}$, on the other
hand, is such that:
\begin{align}
g^{00}  &  =\left(  -1+r_{4}\right)  \mathsf{1}\nonumber\\
g^{01}  &  =g^{10}=r_{0}\mathsf{1}+ r_{1}\sigma_{1}+ r_{2}\sigma_{2}%
+r_{3}\sigma_{3}\label{metric}\\
g^{11}  &  =\left(  1+r_{5}\right)  \mathsf{1}\nonumber
\end{align}
where $\mathsf{1}$ , is the identity and $\sigma_{i}~, i=1,2,3$ are \ the
Pauli matrices.

The target space metric state, which is close to being flat, can be
represented schematically as a density matrix
\begin{equation}
\rho_{\mathrm{grav}}=\int d\,^{5}r\,\,f\left(  r_{\mu}\right)  \left|
g\left(  r_{\mu}\right)  \right\rangle \left\langle g\left(  r_{\mu}\,\right)
\right|  .\, \label{gravdensity}%
\end{equation}
The parameters $r_{\mu}\,\left(  \mu=0,\ldots,5\right)  $ \ are stochastic
with a gaussian distribution $\,f\left(  r_{\mu}\,\right)  $ characterised by
the averages%
\begin{equation}
\label{stoch}\left\langle r_{\mu}\right\rangle =0,\;\left\langle r_{\mu}%
r_{\nu}\right\rangle =\Delta_{\mu}\delta_{\mu\nu}\,.
\end{equation}
The fluctuations experienced by the two entangled neutral mesons will be
assumed to be \emph{independent}.

The above parametrisation has been taken for simplicity and we will also
consider motion to be in the $x$- direction which is natural since the meson
pair moves collinearly in the Center-of-Mass frame.

Space-time deformations of the form (\ref{metric}) and the associated
Hamiltonians (\ref{GenKG}) have been derived in the context of conformal field
theory in earlier works by one of the authors and
collaborators~\cite{Dfoam,szabo2} and details on the relevant derivations will
not be given here. We only mention that the variable $r_{1}$ in particular,
expresses a momentum transfer during the interaction of the (string) matter
state with the D-particle defect. In this sense, the off-diagonal metric
component $g_{01}$ can be represented as
\begin{equation}
g_{01} \sim u_{1} \label{umetric}%
\end{equation}
where $u_{1} = g_{s}\frac{\Delta k}{M_{s}}$ expresses the momentum transfer
along the direction of motion of the matter string (taken here to be the x
direction). In the above equation, $M_{s}/g_{s}$ is the mass of the
D-particle, which for weakly coupled strings with coupling $g_{s}$ is larger
than the string mass scale $M_{s}$. In order to address oscillation phenomena,
induced by D-particles, the fluctuations of each component of the metric
tensor are taken in \cite{bernabeu} to have a $2\times2$ (``flavour'')
structure, as in (\ref{metric}), and not the simple structure (\ref{umetric})
considered in earlier works. In the case of neutral Kaons, which we
concentrate on for concreteness in this section, we use the following notation
for the ``flavours'':
\begin{align}
\left|  K_{L}\right\rangle =\left|  \uparrow\right\rangle ~, \quad\left|
K_{S}\right\rangle =\left|  \downarrow\right\rangle ~.\nonumber
\end{align}
which represent the two physical eigenstates, with masses $m_{1} \equiv m_{L}%
$, $m_{2} \equiv m_{S}$, with
\begin{equation}
\Delta m = m_{L} - m_{S} \sim3.48 \times10^{-15}~\mathrm{GeV}~. \label{deltam}%
\end{equation}
In this way, the stochastic variables $r_{\mu}$ ((\ref{stoch})) in
(\ref{metric}), are linked with the fluctuations of the D-particle recoil
velocity, by representing the latter as:
\begin{equation}
u_{1} \sim r g_{s} \frac{k}{M_{s}}~,
\end{equation}
upon the above-mentioned technicality of considering flavour changes in
addition to the momentum transfer. Specifically, $|r_i | ={\cal O}\left(g_s r k/M_s\right)$.

In this sense, the detailed discussion in the previous session on the
stochastic fluctuations of the recoil velocity about a zero average value,
translates into rewriting (\ref{stoch}) with variances (c.f. (\ref{nonext}))
\begin{equation}
\Delta_{\mu}\sim g_{0}^{2}\frac{t_{0}{}^{2}}{\alpha^{\prime}}\left(
1+\frac{2}{n}\dots\right)  +\mathcal{O}(\sigma^{2})\sim g_{0}^{2}\left(
\frac{p^{0}}{M_{s}}\right)  ^{2}\left(  1+\frac{2}{n}\dots\right)
+\mathcal{O}(\sigma^{2})~,\quad\mu=1,2\label{varmod}%
\end{equation}
where we considered the capture time $t_{0}\sim\alpha^{\prime}p^{0}$, with
$p^{0}$ the energy of the probe, as spanning the essential interaction time
with the D-particle of the initial entangled meson state. In this way we extrapolate the result (\ref{gaussian2}) to times smaller than $\sqrt{\alpha^{\prime}}$. This is
acceptable, as long as such times are \emph{finite}. From a conformal field
theory point of view, this means that we consider the world-sheet scaling
parameter $1/\varepsilon^{2}\sim\mathrm{ln}(L/a)^{2}\sim t_{0}\ll\sqrt
{\alpha^{\prime}}$ for probe energies $p^{0}\ll M_{s}$.

We next note that the Hamiltonian interaction terms
\begin{equation}
\widehat{H_{I}}=-\left(  {r_{1}\sigma_{1}+r_{2}\sigma_{2}}\right)  \widehat
{k}\label{inter}%
\end{equation}
are the leading order contribution in the small parameters $r_{\mu}$ in the
Hamiltonian $H$ (\ref{GenKG}), since the corresponding variances $\sqrt
{\Delta_{\mu}}$ are small. The term (\ref{inter}), has been used in
\cite{bernabeu} as a \emph{perturbation} in the framework of non-degenerate
perturbation theory, in order to derive the \textquotedblleft
gravitationally-dressed\textquotedblright\ initial entangled meson states,
immediately after the $\phi$ decay. The result is:
\begin{align}
&  \left\vert {k,\uparrow}\right\rangle _{QG}^{\left(  1\right)  }\left\vert
{-k,\downarrow}\right\rangle _{QG}^{\left(  2\right)  }-\left\vert
{k,\downarrow}\right\rangle _{QG}^{\left(  1\right)  }\left\vert {-k,\uparrow
}\right\rangle _{QG}^{\left(  2\right)  }=\left\vert {k,\uparrow}\right\rangle
^{\left(  1\right)  }\left\vert {-k,\downarrow}\right\rangle ^{\left(
2\right)  }-\left\vert {k,\downarrow}\right\rangle ^{\left(  1\right)
}\left\vert {-k,\uparrow}\right\rangle ^{\left(  2\right)  }\nonumber\\
&  +\left\vert {k,\downarrow}\right\rangle ^{\left(  1\right)  }\left\vert
{-k,\downarrow}\right\rangle ^{\left(  2\right)  }\left(  {\beta^{\left(
1\right)  }-\beta^{\left(  2\right)  }}\right)  +\left\vert {k,\uparrow
}\right\rangle ^{\left(  1\right)  }\left\vert {-k,\uparrow}\right\rangle
^{\left(  2\right)  }\left(  {\alpha^{\left(  2\right)  }-\alpha^{\left(
1\right)  }}\right)  \nonumber\\
&  +\beta^{\left(  1\right)  }\alpha^{\left(  2\right)  }\left\vert
{k,\downarrow}\right\rangle ^{\left(  1\right)  }\left\vert {-k,\uparrow
}\right\rangle ^{\left(  2\right)  }-\alpha^{\left(  1\right)  }\beta^{\left(
2\right)  }\left\vert {k,\uparrow}\right\rangle ^{\left(  1\right)
}\left\vert {-k,\downarrow}\right\rangle ^{\left(  2\right)  }\label{entangl}%
\end{align}
where
\begin{equation}
\alpha^{\left(  i\right)  }=\frac{^{\left(  i\right)  }\left\langle
\uparrow,k^{\left(  i\right)  }\right\vert \widehat{H_{I}}\left\vert
k^{\left(  i\right)  },\downarrow\right\rangle ^{\left(  i\right)  }}%
{E_{2}-E_{1}}~,\quad\beta^{\left(  i\right)  }=\frac{^{\left(  i\right)
}\left\langle \downarrow,k^{\left(  i\right)  }\right\vert \widehat{H_{I}%
}\left\vert k^{\left(  i\right)  },\uparrow\right\rangle ^{\left(  i\right)
}}{E_{1}-E_{2}}~,~\quad i=1,2\label{qgpert2}%
\end{equation}
where the index $(i)$ runs over meson species (\textquotedblleft
flavours\textquotedblright) ($1\rightarrow K_{L},~2\rightarrow K_{S}$). The
reader should notice that the terms proportional to $\left(  {\alpha^{\left(
2\right)  }-\alpha^{\left(  1\right)  }}\right)  $ and $\left(  {\beta
^{\left(  1\right)  }-\beta^{\left(  2\right)  }}\right)  $ in (\ref{entangl})
generate $\omega$-like effects. We concentrate here for brevity and
concreteness in the strangeness conserving case of the $\mathbb{\omega}%
$-effect in the initial decay of the $\phi$ meson~\cite{bernabeu1}, which
corresponds to $r_{i}\propto\delta_{i2}$. We should mention, however, that in
general quantum gravity does not have to conserve this quantum number, and in
fact strangeness-violating $\mathbb{\omega}$-like terms are generated in this
problem through time evolution~\cite{bernabeu}.

We next remark that on averaging the density matrix (c.f. (\ref{gravdensity})) over the random variables
$r_{i}$, which are treated as independent variables between the two meson
particles of the initial state (\ref{entangl}), we observe that only terms of
order $|\mathbb{\omega}|^{2}$ will survive, with the order of $|\mathbb{\omega
}|^{2}$ being
\begin{equation}
|\mathbb{\omega}|^{2}=\widetilde{\sum}_{(1),(2)}\left(  \mathcal{O}\left(
\frac{1}{(E_{1}-E_{2})^2}(\langle\downarrow,k|H_{I}|k,\uparrow\rangle
)^{2}\right)  \right)  =\widetilde{\sum}_{(1),(2)}\left(  \mathcal{O}\left(
\frac{\Delta_{2}k^{2}}{(E_{1}-E_{2})^{2}}\right)  \right)  \sim\widetilde
{\sum}_{(1),(2)}\left(  \frac{\Delta_{2}k^{2}}{(m_{1}-m_{2})^{2}}\right)
\label{omegaorder}%
\end{equation}
for the physically interesting case of non-relativistic Kaons in $\phi$
factories, in which the momenta are of order of the rest energies. The
notation $\widetilde{\sum}_{(1),(2)}\left(  \dots\right)  $ above indicates
that one considers the sum of the variances $\Delta_{2}$ over the
two meson states $1$, $2$ as defined above.

The variances in our model of D-foam, which are due to quantum fluctuations of
the recoil velocity variables about the zero average (dictated by the imposed
requirement on Lorentz invariance of the string vacuum) are given by
(\ref{varmod}), with $p^{0}\sim m_{i}$ the \emph{energy} of the corresponding
individual (non-relativistic) meson state $(i)$, $i=1,2$, in the initial entangled state
(\ref{entangl}). It is important to notice that, on taking the sum of
the variance $\Delta_{2}$ over the mesons (1) and (2), the terms
proportional to the dispersion $\sigma^{2}$ in the initial recoil velocity
$u_{0}$ Gaussian distribution in (\ref{nonext}) give a contribution of order $30\sigma^2$, since $\sigma^{2}$ is assumed universal among particle species. This is a parameter that depends on the details of the foam. As already mentioned in the previous section, one may assume models in which $\sigma^2 \ll (g_0\sqrt{\alpha '} p^0)^2$. In this sense, one is left with the contributions from the first
term of the right-hand-side of (\ref{nonext}), and thus we obtain the
following estimate for the square of the amplitude of the (complex) $\mathbb{\omega}$-parameter:
\begin{equation}
|\omega|^{2}\sim g_{0}^{2}\frac{\left(  m_{1}^{2} + m_{2}^{2}\right)  }%
{M_{s}^{2}}\frac{k^{2}}{(m_{1}-m_{2})^{2}}\left(  1+\frac{2}{n}\dots\right)~,
\quad M_{s}/g_{0}\equiv M_{P}~,\label{final}%
\end{equation}
where $M_{P} = M_{s}/g_{0}$ is the (average) D-particle
mass, as already mentioned, representing the (average) quantum gravity scale, which may be taken to be the four-dimensional Planck scale. In the modern version of string theory, $M_{s}$ is arbitrary and can be as low as a few TeV, but in order to have
phenomenologically correct string models with large extra dimensions one also
has to have in such cases very weak string couplings $g_{0}$, such that even
in such cases of low $M_{s}$, the D-particle mass $M_{s}/g_{0}$ is always
close to the Planck scale $10^{19}$ GeV. But of course one has to keep an open
mind about ways out of this pattern, especially in view of the string landscape.

The result (\ref{final}), implies, for neutral Kaons in a $\phi$ factory, for
which (\ref{deltam}) is valid, the estimate $|\mathbb{\omega}| ={\cal O}\left( 10^{-5}\right)$,
which, in the sensitive $\eta^{+-}$ bi-pion decay channel, leads to effects enhanced by three
orders of magnitude, as a result of the fact that the $|\mathbb{\omega}|$
effect always appears in the corresponding observables~\cite{bernabeu1} in the
form $|\mathbb{\omega}|/|\eta^{+-}|$, and the CP-violating parameter
$|\eta^{+-}|\sim10^{-3}$. At present, this value is still some two
orders of magnitude away from current bounds of the $\mathbb{\omega}$-effect
by the KLOE collaboration at DA$\Phi$NE~\cite{dafne}, giving $|\omega| < 10^{-3}$
but it is within the projected sensitivity of the proposed upgrades.

The above estimate is valid for non-relativistic meson states, where each meson has been treated as a structureless entity when considering its interaction with the D-foam.
The inclusion of details of its strongly-interacting substructure may affect the results.
To consider relativistic meson states, the non-relativistic quantum mechanics formalism
leading to (\ref{final}) in the Kaon systems should strictly be replaced by an
appropriate relativistic treatment.
The major
difference in this case is the form of the Hamiltonian, which stems from the
expansion of the Dirac Hamiltonian for momenta $k\gg m_{i}$, $m_{i}$ the
masses. The quantities $E_{i}\sim k+\frac{m_{i}^{2}}{2k},~i=1,2$ (due to
momentum conservation, assumed on average), and the capture times $t_{c}%
\sim\alpha^{\prime}E_{i}$.
Nevertheless, if one extrapolates naively the above results for the
$|\mathbb{\omega}|^{2}$ in such a case, one arrives at the conclusion that
for the validity of leading-order perturbation theory for the interaction with the D-foam,  one should consider momenta of order $k \sim 30$ GeV. Otherwise higher-order corrections become important.
This completes our discussion on the
estimates of the $\mathbb{\omega}$-effect in the initial entangled state of
two mesons in a meson factory. As discussed in \cite{bernabeu},
$\mathbb{\omega}$-like terms can also be generated due to the time evolution.
One can apply similar estimates for this case too. We shall not do so in this work.

\section{Conclusions and Outlook \label{sec:7}}

This work examines the r\^ole of quantum string fluctuations (which can give
rise to a non-commutative space-time geometry at string scales) on the velocity distribution of
\ D-particles, within a specific kind of foam in string theory.
Our Gaussian modeling of the recoil velocity is found to be
robust to these fluctuations.

In this way we have managed to give a rather rigorous estimation (modulo
strong interaction effects) of the $\mathbb{\omega}$-effect
in entangled states of mesons, which has been compared to previous naive estimates, based on dimensional analysis. The effect is
smaller by roughly two orders of magnitude from the current upper bounds
set by the KLOE collaboration at DA$\Phi$NE. This
makes the detectability of the effect in future meson factories a major experimental challenge.

Admittedly, our approach in this paper is based on bosonic string theory
which is not the most relevant phenomenologically. World-sheet supersymmetric
strings do not lead to a closed-form resummation of the leading divergencies of the pinched surfaces, since the latter cancel out~\cite{szabo2,mav2}, and the remaining terms are hard to
cast in a closed form.
However, despite this apparent technical difficulty,  the general
conclusions drawn from the current work, as far as fuzzyness of the string coupling and the target space time are concerned, are likely to be robust, since they depend on the form of vertex
operators for (recoil) zero modes of D-particles (provided of course the theories are
restricted to those admitting D-particles).
In this sense, our approach here opens up the possibility of
extending the analyses of D-particle foam to non-extensive statistics. The
latter as a class is ubiquitous but, as far as we are aware, the possibility
of non-extensive statistics in D-particle recoil has not hitherto been raised.
Moreover, the mechanism makes contact with some non-trivial features of
string theory.

The r\^ole of such non-extensivity on matter propagation in
D-particle foam, matter number distribution functions and (supercritical) string
cosmology~\cite{lmn} will be addressed in future publications.
This is an important aspect of the formalism discussed here, since it may have a profound influence on the dark matter (and dark energy) distributions in a Universe with D-particle foam, which may have phenomenological consequences, as far as
constraints on, say, supersymmetric particle physics models are concerned.
Indeed, as we have discussed above, the interaction of D-particles with matter leads to local distortions of the neighboring space-time (\ref{finsler}), which depend on both the string coupling $g_s$, through the D-particle masses $M_s/g_s$, and the recoil velocity of the D-particle (i.e. the momentum transfer
of the string matter) $u_i$, which stochastically fluctuates upon summing up higher-genus
world-sheet topologies. In view of our discussion in this work, both these quantities can be ``fuzzy'', leading to stochastic fluctuations on the space-time metric, on which
string matter lives. These fluctuations are up and above any statistical fluctuations in populations of D-particles that characterise the D-particle foam models.

The presence of such fluctuations affect important cosmological
quantities that are directly relevant to the (supercritical) Universe budget, such as thermal supersymmetric dark matter relic densities, through appropriate modifications of the
relevant thermodynamic equations. Hence, the relevant astro-particle physics constraints on supersymmetric models~\cite{lmn} are also modified. As we have noted above, however, the issue as to whether the fuzzyness of the string coupling and consequently the non-extensivity of the D-particle foam can lead to observable signatures in Cosmology or astro-particle physics in general, remains to be seen.
We hope to be able to report in a more detailed form on these phenomenological issues in the near future.

\bigskip

\section*{Acknowledgements}

This work is partially supported by the European Union through the FP6 Marie
Curie Research and Training Network \emph{UniverseNet} (MRTN-CT-2006-035863).

\section*{Appendix A: Recoil and Leading Divergences in the World-Sheet Genus Expansion}

In this appendix we shall show how the leading $(\log\delta)^2$ modular
divergences which appear in \eqn{mixing} can be removed by invoking an
appropriate Ward identity for the fundamental string fields of the matrix
$\sigma$-model. As we shall show, this is equivalent to imposing \emph{momentum
conservation} for scattering processes in the matrix D-brane background.
We shall be brief in our discussion, restricting oursleves on the main results,
of relevance to our discussion in this work.
The material in this section is taken from \cite{szabo}, where we refer the
interested reader for further details.

\begin{figure}[ht]
\centering
\includegraphics[width=7.5cm]{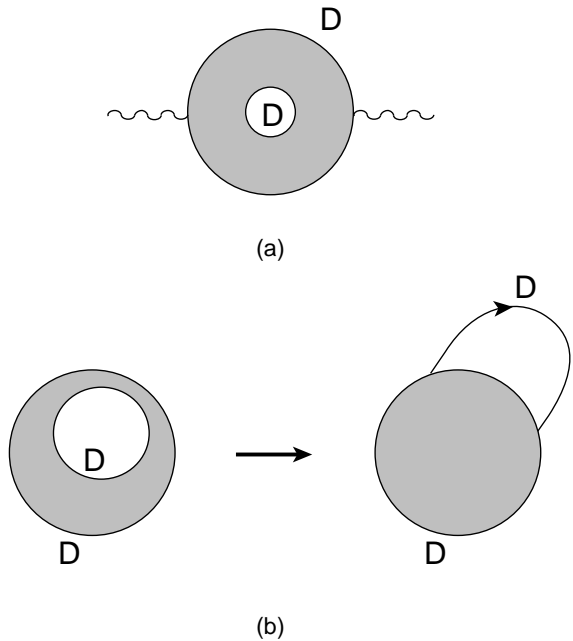}
\caption{(a) World-sheet annulus diagram for the
leading
quantum correction to the propagation of a string state $V$ in a D-brane
background, and (b) the pinched annulus configuration which is the
dominant divergent contribution to the quantum recoil.}
\label{fig:annulus}\end{figure}

As shown in that reference, conformal invariance
requires absorbing such singularities into renormalized quantities at lower
genera, leading to a generalized version of the Fischler-Susskind mechanism
\cite{fs}. Such degenerate Riemann surfaces involve a string propagator over
thin long worldsheet strips of thickness $\delta \to 0$ that are attached to a
disc. These strips can be thought of as two-dimensional quantum gravity
wormholes. Consider first the resummation of one-loop worldsheets, i.e. those
with an annular topology, in the pinched approximation (c.f. fig.~\ref{fig:annulus}). String
propagation on such a worldsheet can be described formally by adding bilocal
worldsheet operators $\cal B$, which in the present case are
defined by~\cite{szabo}
\bea
{\cal
B}(\omega,\omega')=\sum_{a,b,c,d}\,\oint_{\partial\Sigma}\oint_{\partial
\Sigma'}V_{ab}^i(x;\omega)~\frac{G_{ij}^{ab;cd}(
\omega,\omega')}{L_0-1}~V_{cd}^j(x;\omega')
\label{bilocal}\eea
where $G_{ij}$ denotes the Zamolodchikov metric in string theory space, that is the two-point correlation
function of the recoil vertex operators:
\begin{eqnarray}
\oint_{\partial\Sigma_h}V^i_{ab}(x;\omega)\equiv\lim_{\varepsilon\to0^+}\frac
{ig_s}{2\pi\alpha'}\sum_{k=0}^h\int_0^1ds_k~\e^{-i\omega
X^0(s_k)}\,\Theta(X^0(s_k);\varepsilon)
\,\bar\xi_a(s_k-\varepsilon)\xi_b(s_k)\,\frac d{ds_k}X^i(s_k)~,
\label{newvertexops}
\end{eqnarray}
and $L_0$
denotes the usual Virasoro generator. The variable $\omega$ is a Fourier variable, appearing in the Fourier transform
\begin{equation}
\breve{Y}_i^{ab}(\omega)=\lim_{\varepsilon \to 0^+}\int_0^\infty dt~\e^{i\omega
t}\,Y_i^{ab}(X^0;\varepsilon)
\label{ftcollcoord}
\end{equation}
of the collective coordinates $Y_i^{ab}(X^0; \varepsilon)$ of the (group of) D-particles~\cite{kogan,szabo} \begin{equation}
    Y_i^{ab} (X^0 ; \varepsilon ) = \sqrt{\alpha '} Y_i \varepsilon + u_i^{ab} X^0~.
    \label{collcoord}
    \end{equation}
where $Y_i^{ab}, u_j^{ab}$ denote the collective coordinates and recoil velocities of the group of $N$
D-particles, with the indices $a,b = 1, \dots N$.
The reader should also note that, as explained in \cite{szabo}, the range of the variable $\omega \in (0, \infty)$, as a result of the impulse approximation of recoil, which dictates that the recoil vertex operators are non trivial only for target times after some definite time (taken to be zero for concreteness in the example discussed here (\ref{fullrec})).
The operator insertion $(L_0-1)^{-1}$ in
\eqn{bilocal} represents the string propagator $\triangle_s$ on the thin strip
of the pinched annulus.

Inserting a complete set of intermediate string states ${\cal E}_I$, we can
rewrite (\ref{bilocal}) as an integral over the variable $q \equiv \e^{2 \pi i
\tau}$, where $\tau$ is the complex modular parameter characterizing the
worldsheet strip. The string propagator over the strip then reads
\be
\triangle_s(z,z')\,=\,\sum_I\int dq~q^{\Delta_I-1}\,\Bigl\{{\cal
E}_I(z)\otimes({\rm ghosts})\otimes{\cal E}_I(z')\Bigr\}_{\Sigma\#\Sigma'}
\label{props}\ee
where $\Delta_I$ are the conformal dimensions of the states ${\cal E}_I$. The
sum in (\ref{props}) is over all states which propagate along the long thin
strip connecting the discs $\Sigma$ and $\Sigma'$ (in the degenerating annulus
handle case of interest here, $\Sigma'=\Sigma$). As indicated in (\ref{props}),
the sum over states must include ghosts, whose central charge cancels that of
the worldsheet matter fields in any critical string model.

In \eqn{props} we have assumed that the Virasoro operator $L_0$ can be
diagonalized in the basis of string states with eigenvalues their conformal
dimensions $\Delta_I$, i.e.
\be
L_0|{\cal E}_I\rangle=\Delta_I|{\cal E}_I\rangle~~~~~~,~~~~~~q^{L_0-1}|{\cal
E}_I\rangle=q^{\Delta_I-1}|{\cal E}_I\rangle
\label{diagonal}\ee
However, this simple diagonalization fails~\cite{kogan} in the presence of the logarithmic
pair of operators (\ref{newvertexops}), due to the non-trivial mixing between $C$
and $D$ in the Jordan cell of $L_0$. Generally, states with
$\Delta_I=0$ may lead to extra logarithmic divergences in (\ref{props}),
because such states make contributions to the integral of the form $\int
dq/q\sim\log\delta$, in the limit $q \sim \delta \rightarrow 0$
representing a long thin strip of thickness $\delta$. We assume that such
states are discrete in the space of all string states, i.e. that they are
separated from other states by a gap. In that case, there are factorizable
logarithmic divergences in (\ref{props}) which depend on the background
surfaces $\Sigma$ and $\Sigma'$. These are precisely the states corresponding
to the logarithmic recoil operators (\ref{Depsilonop}) and \eqn{Cepsilonop},
with vanishing conformal dimension $\Delta_\varepsilon = -\varepsilon^2/2$  as $\varepsilon \rightarrow
0^+$.

In the case of mixed logarithmic states, the pinched world-sheet topologies
of fig.~\ref{fig:dbranes}
are characterized by divergences of a double logarithmic type which arise from the
form of the string propagator in (\ref{bilocal}) in the presence of generic
logarithmic operators $C$ and $D$,
\bea
\int dq~q^{\Delta_\varepsilon-1}\,\langle C,D|\begin{pmatrix} 1&\log
q\cr0&1\cr \end{pmatrix}|C,D\rangle
\label{CDprop}\eea
As shown in \cite{kogan}, the mixing between $C$ and $D$ states along
degenerate handles leads formally to divergent string propagators in physical
amplitudes, whose integrations have leading divergences of the form
\bea
\int\frac{dq}q~\log q\int d^2z~D(z;\varepsilon)\int
d^2z'~C(z';\varepsilon)\simeq(\log\delta)^2
\int d^2z~D(z;\varepsilon)\int d^2z'~C(z';\varepsilon)
\label{mixing}\eea
These $(\log\delta)^2$ divergences can be cancelled by imposing momentum
conservation in the scattering process of the light string states off the
D-brane background \cite{szabo}.

We notice at this stage that for brevity we have discussed in this work
the simplistic case of a single D-particle interacting with an (open) string state.
However, as a result of flux conservation, isolated D-particles do not exist, as we
have mentioned above, and hence in realistic situations one deals with groups of $N$ (with $N$ varying)
D-particles, interacting among themselves with the exchange of stretched flux-carrying strings.
Such groups of $N$ particles can be represented by a non-Abelian Wilson loop operator for the gauge group $SU(N)$. Within the framework of the auxiliary field representation of the Wilson loop
operator~\cite{szabo},
the effective abelianization of the matrix $\sigma$-model leads to a
relatively straightforward generalization of the proof of the cancelation of modular
infinities by introducing recoil, as we now demonstrate.

The pertinent bilocal term induced by (\ref{mixing}), which exponentiates upon
summing over pinched topologies, can be written as a local worldsheet effective
action using the wormhole parameters $[\rho_{C,D}]_i^{ab}$
to give~\cite{szabo}
\bea
\e^{\Delta S^{CD}}&=&\lim_{\varepsilon\to0^+}\int d\rho_C~d\rho_D~
\exp\left[\sum_{a,b=1}^N\left(-\frac1{2g_s^2(\log\delta)^2}\,G^{LM}
\sum_{c,d=1}^NG_{ab;cd}^{ij}\,[\rho_L]_i^{ab}[\rho_M]_j^{cd}\right.\right.
\nn\\& &~~~~~~~~~~~~~~~+\frac{ig_s[\rho_C]_i^{ab}}{2\pi\alpha'}
\int_0^1ds~C(X^0(s);\varepsilon)\,
\bar\xi_a(s-\varepsilon)\xi_b(s)\,\frac d{ds}x^i(s)\nn\\& &~~~~~~~~~~~~~~~
\left.\left.+\frac{ig_s[\rho_D]_i^{ab}}{2\pi\alpha'}\int_0^1ds~D(X^0(s);
\varepsilon)\,\bar\xi_a(s-\varepsilon)\xi_b(s)\,\frac d{ds}x^i(s)\right)\right]
\label{wlocal2}\eea
Here we have for simplicity considered only the zero frequency modes of the
fields involved with respect to the Fourier transformations defined in \cite{szabo}. They will be sufficient to describe the relevant
cancelations. In \eqn{wlocal2} the (dimensionless) moduli space metric
$G^{LM}$ (where $L,M=C,D$) is an appropriate off-diagonal $2\times2$ matrix.
which
is required to reproduce the initial bilocal operator with the $CD$-mixing of
the logarithmic operators. This off-diagonal metric includes all the
appropriate normalization factors ${\cal N}_L$ for the zero mode states. These
factors are essentially the inverse of the $CD$ two-point function,
which is finite.

We consider the propagation of two (closed string) matter tachyon states
$T_{1,2}=\e^{i(k_{1,2})_ix^i}$ in the background of (\ref{wlocal2}) at tree
level. In what follows the effects of the $C$ operator are sub-leading and can
be ignored. Then, we are interested in the amplitude
\bea
{\cal A}_{CD}&\equiv&\left\langle~\left\langle\!\!\left\langle\sum_{c'=1}^N
\bar\xi_{c'}(0)\,T_1T_2~\e^{\Delta
S^{CD}}\,\xi_{c'}(1)\right\rangle\!\!\right\rangle~\right\rangle_0\nonumber\\
&=&\lim_{\varepsilon\to0^+}\sum_{c'=1}^N\int d\rho_C~d\rho_D~\int
Dx~D\bar\xi~D\xi~\bar\xi_{c'}(0)\nonumber\\&
&\times\exp\left(-N^2S_0[x]-\sum_{c=1}^N\int_0^1ds~
\bar\xi_c(s-\varepsilon)\frac d{ds}\xi_c(s)\right)\nonumber\\&
&\times\,T_1[x]\,T_2[x]~\exp\left[\sum_{a,b=1}^N
\left(-\frac1{2g_s^2(\log\delta)^2}\,G^{LM}
\sum_{c,d=1}^NG_{ab;cd}^{ij}\,[\rho_L]_i^{ab}[\rho_M]_j^{cd}\right.\right.
\nn\\& &\left.\left.+\frac{ig_s[\rho_D]_i^{ab}}{2\pi\alpha'}
\int_0^1ds~D(X^0(s);\varepsilon)\,
\bar\xi_a(s-\varepsilon)\xi_b(s)\,\frac d{ds}x^i(s)\right)\right]\xi_{c'}(1)+\dots
\label{dsv1v2}\eea
where $\dots$ represent sub-leading terms. The scaling property
\eqn{scale2} of the logarithmic operators must be taken into account.
Under a scale transformation on the worldsheet the $C$
operator emerges from $D$ due to mixing with a scale-dependent coefficient
$\sqrt{\alpha'}\,t$. This will contribute to the scaling infinities we are
considering here.

The composite $D$ operator insertion in (\ref{dsv1v2}) needs to be
normal-ordered on the disc. Normal ordering in the present case amounts to
subtracting scaling infinities originating from divergent contributions of
$D(X^0(s);\varepsilon)$ as $\varepsilon \rightarrow 0^+$. To determine these
infinities, we first note that the one-point function of the composite $D$
operators, computed with respect to the free $\sigma$-model and auxiliary
field actions, can be written as
\bea
& &\left\langle~\left\langle\!\!\left\langle\sum_{c'=1}^N\bar\xi_{c'}(0)
\exp\left(\sum_{a,b=1}^N\frac{ig_s[\rho_D]_i^{ab}}{2\pi\alpha'}
\int_0^1ds~D(X^0(s);\varepsilon)\,\bar\xi_a(s-\varepsilon)\xi_b(s)\,\frac
d{ds}x^i(s)\right)\xi_{c'}(1)\right\rangle\!\!\right\rangle~
\right\rangle_0\nn\\&
&=~\left\langle\!\!\left\langle\sum_{c'=1}^N\bar\xi_{c'}(0)\exp\left(-
\sum_{a,b,c,d}\frac{g_s^2[\rho_D]_i^{ab}[\rho_D]_j^{cd}}{2(2\pi\alpha')^2}
\int_0^1ds~ds'~\left\langle
D(X^0(s);\varepsilon)\,D(X^0(s');\varepsilon)\right\rangle_0\right.\right.
\right.\nn\\& &~~~~~~~~~~~~~~~~~~~~~~~~~~~~~~
\Biggl.\Biggl.\Biggl.\times\,\bar\xi_a(s-\varepsilon)
\xi_b(s)\bar\xi_c(s'-\varepsilon)\xi_d(s')\left\langle\mbox{$\frac
d{ds}$}\,x^i(s)\,\mbox{$\frac
d{ds'}$}\,x^j(s')\right\rangle_0\Biggr)\xi_{c'}(1)\Biggr\rangle\!\!\Biggr
\rangle\nn\\&
&=~\exp\left(-\sum_{a,b=1}^N\frac{g_s^2[\rho_D]_i^{ab}[\rho_D]_j^{ba}}
{2(2\pi\alpha')^2}\int_0^1ds~ds'~\left\langle
D(X^0(s);\varepsilon)\,D(X^0(s');\varepsilon)\right\rangle_0\left\langle\mbox{$\frac
d{ds}$}\,x^i(s)\mbox{$\frac d{ds'}$}\,x^j(s')\right\rangle_0\right)\nn\\& &~~~~
\label{two-point}\eea
where we have used Wick's theorem. The second equality in \eqn{two-point}
follows after removing ambiguous $\Theta(\varepsilon)$ type terms from the Wick
expansion in the auxiliary fields using the renormalization scheme described
in appendix B of \cite{szabo}. One finds that this procedure has the overall effect of
replacing the product of auxiliary fields in the first equality in
\eqn{two-point} by the delta-functions $\delta_{ad}\delta_{bc}$.

In what follows we shall ignore, for simplicity, the basic divergences that
come from the fundamental string propagator in \eqn{two-point}. Such
divergences will appear globally in all correlators below and will not affect
the final result. As a consequence of the logarithmic algebra and
the scale transformation \eqn{scale2}, there are
leading (scaling) divergences in \eqn{two-point} for
$\varepsilon\rightarrow 0^+$ which behave as
\bea
g_s^2b\alpha'^{-1/2}t~\tr\,[\rho_D]_i[\rho_D]^i
\label{counterterm}\eea
Thus, normal ordering of the $D$ operator amounts to adding a term of opposite
sign to \eqn{counterterm} into the argument of the exponential in
(\ref{dsv1v2}) in order to cancel such divergences.

Let us now introduce a complete set of states $|{\cal E}_I\rangle$ into the
two-point function of string matter fields on the disc,
\be
\langle T_1T_2\rangle_0=\sum_I|{\cal N}_I|^2\,\langle T_1|{\cal
E}_I\rangle_0\,\langle{\cal E}_I|T_2\rangle_0
\label{completeset}\ee
where ${\cal N}_I$ is a normalization factor for the fundamental string states
(determined by the Zamolodchikov metric). Taking into account the effects of
the $C$ operator included in $D$ under the finite-size scaling (\ref{fsscaling}), we
see that the leading divergent contributions to (\ref{completeset}) are of the
form
\be
\langle T_1T_2\rangle_0\simeq-\sqrt{\alpha'}\,t\,\langle
T_1|C\rangle_0\,\langle
C|T_2\rangle_0+\dots
\label{vc}\ee
where we have used (\ref{epscutoff}), as well as the fact (c.f. (\ref{canep1}), (\ref{canep2}) and (\ref{canep3})) that the Zamolodchikov metric in the $C,D$ basis behaves as~\cite{kogan}:
\bea
G_{CC}\sim\varepsilon^2~~~~~~,~~~~~~G_{DD}\sim\varepsilon^{-2}~~~~~~,~~~~~~
G_{CD}=G_{DC}\sim{\rm const.}~.
\label{ZCDscale}\eea
We now notice that the
$C$ deformation vertex operator plays the role of the Goldstone mode for the
translation symmetry of the fundamental string coordinates $x^i$, and as such
we can apply the corresponding Ward identity in the matrix $\sigma$-model path
integral to represent the action of the $C$ deformation on physical states by
$-i\delta/\delta x^i$ \cite{kogan,coll}. The leading contribution to
\eqn{completeset} can thus be exponentiated to yield
\bea
\langle T_1T_2\rangle_0&\simeq&\lim_{\varepsilon\to0^+}\sum_{c'=1}^N\int
Dx~D\bar\xi~D\xi~\bar\xi_{c'}(0)~\exp\left(-N^2S_0[x]-\sum_{c=1}^N\int_0^1ds~
\bar\xi_c(s-\varepsilon)\frac d{ds}\xi_c(s)\right)\nn\\&
&~~~~~~\times\,T_1[x]\exp\left(-\frac{g_s^2\sqrt{\alpha'}\,t}2
\sum_{a,b=1}^N\int_0^1ds~ds'~\bar
\xi_a(s-\varepsilon)\xi_b(s)\bar\xi_b(s'-\varepsilon)\xi_a(s')\right.\nn\\&
&~~~~~~\left.\times\,\frac{\buildrel\leftarrow\over\delta}{\delta
x_i(s)}\frac{\buildrel\rightarrow\over\delta}{\delta
x^i(s')}\right)T_2[x]~\xi_{c'}(1)
\label{expon}\eea
where we have used the on-shell condition $T_j(\frac\delta{\delta
x_i}\frac\delta{\delta x^i})T_k=0$ for the tachyon fields. \eqn{expon}
expresses the non-abelian version of the Ward identity in the presence of
logarithmic deformations.

Using \eqn{counterterm}, \eqn{expon} and normalizing the parameters of the
logarithmic conformal algebra appropriately, it follows that (\ref{dsv1v2}) can
be written as~\cite{szabo}
\bea
{\cal A}_{CD} & = & \lim_{\varepsilon\to0^+}
\sum_{c'=1}^N\int d\rho_C~d\rho_D~\int Dx~D\bar\xi~D\xi~\bar\xi_{c'}(0)\nn\\&
&\times\exp\left(-N^2S_0[x]-\sum_{c=1}^N\int_0^1ds~
\bar\xi_c(s-\varepsilon)\frac d{ds}\xi_c(s)\right)\nn\\&
&\times\,T_1[x]\exp\left[\sum_{a,b=1}^N
\left\{-\frac1{2g_s^2(\log\delta)^2}\,G^{LM}
\sum_{c,d=1}^NG_{ab;cd}^{ij}\,[\rho_L]_i^{ab}[\rho_M]_j^{cd}\right.\right.
\nn\\& &-\frac{g_s^2\alpha'^{-1/2}t}2\,\eta^{ij}\left([\rho_D]_i^{ab}-\frac{i
\sqrt{\alpha'}}{g_s}\int_0^1ds~\bar\xi_a(s-\varepsilon)\xi_b(s)
\,\frac{\buildrel\leftrightarrow\over\delta}{\delta x_i(s)}\right)\nn\\&
&~~~~~~\left.\left.\times\left([\rho_D]_j^{ba}-\frac{i\sqrt{\alpha'}}{g_s}
\int_0^1ds~\bar\xi_b(s-
\varepsilon)\xi_a(s)\,\frac{\buildrel\leftrightarrow\over\delta}{\delta
x_j(s)}\right)\right\}\right]T_2[x]~\xi_{c'}(1)+\dots
\label{cdsq2}\eea
{}From (\ref{cdsq2}) it follows that the limit $t\to\infty$ localizes the
worldsheet wormhole parameter integrations with delta-function support
\bea
\prod_{a,b=1}^N\,\prod_{i=1}^9\delta\left(\mbox{$[\rho_D]_i^{ab}-\mbox{$
\frac{\sqrt{\alpha'}}{g_s}$}\,
\,(k_1+k_2)_i\,\int_0^1ds~\bar\xi_a(s-\varepsilon)\xi_b(s)$}\right)
\label{dfunc}\eea
where $(k_{1,2})_{i}$ are the momenta of the closed string matter states. This
result shows that the leading modular divergences in the genus expansion are
cancelled by the scattering of (closed) string states off the matrix D-brane
background. Upon rescaling $\rho_D$ by $g_s^2$, averaging over the auxiliary
boundary fields, and incorporating \eqn{dfunc} as an effective shift in the
velocity recoil operator (c.f. (\ref{qcoupling})): \bea
\breve{Y}_i^{ab}(\omega)~\to~\widehat{\cal
Y}_i^{ab}(\omega)=\breve{Y}_i^{ab}(\omega)+g_s\,\sqrt{\log\delta}~
\breve{\rho}_i^{ab}(\omega)
\label{quantumcoords}\eea
viewed as position operator in a co-moving target space frame,
we can identify this
renormalization as fixing the velocity matrix
\bea
U_i^{ab}=-\sqrt{\alpha'}\,g_s\,(k_1+k_2)_i\,\delta^{ab}
\label{velocitydiag}\eea
of the fat brane background. Thus momentum conservation for the D-brane
dynamics guarantees conformal invariance of the matrix $\sigma$-model as far as
leading divergences are concerned.

\vfill

\section*{Appendix B: Non-Extensive (Tsallis-type) statistics}

In this Appendix we review the analysis of \cite{beck} in constructing classes
of stochastic differential equations with fluctuating frictional forces which
generate dynamics described by non-extensive statistics in the sense of
Tsallis~\cite{tsallis}.

Our starting point is the linear Langevin equation describing the dynamics of
a Brownian particle:
\begin{equation}
{\dot u} = -\gamma u + \sigma L(t) \label{langevinbeck}%
\end{equation}
where $L(t)$ is a Gaussian white noise, $\gamma> 0$ is a friction constant
coefficient, and the (real) parameter $\sigma$ described the strength of the noise.

We can take the probability density of the velocity field $u$ to be Gaussian
with average $\langle u \rangle= 0$ and variance $\langle v^{2} \rangle=
\beta^{-1}, $ with $\beta= \frac{\gamma}{\sigma^{2}}$, which plays the r\^ole
of an inverse temperature of the Brownian particle, taken for simplicity to be
of unit mass. To get Tsallis statistics, we have to allow for the parameter
$\beta$ to fluctuate, which can be achieved by allowing in whole generality
the parameters $\gamma$ and $\beta$ to fluctuate as well.

The case studied in \cite{beck}, and used in our work in this article, is the
one in which the parameter $\beta$ is $\chi^{2}$ distributed with degree $n$,
i.e. the probability density of $\beta$ is given by:
\begin{equation}
p(\beta) = \frac{1}{\Gamma(\frac{n}{2})}\left(  \frac{n}{2\beta_{0}}\right)
^{n/2}\beta^{n/2-1}\mathrm{exp}\left(  -\frac{n\beta}{2\beta_{0}}\right)  ~,
\label{chidistribution}%
\end{equation}
with $\beta_{0}$ a constant, playing the r\^ole of the average of the $\beta$ fluctuations.

There are many examples which lead to the distribution of the form
(\ref{chidistribution}), for example~\cite{beck} if $\beta$ is given by a sum
of squares of independent Gaussian random variables $X_{i}~, i=1, \dots n$:
\begin{equation}
\beta= \sum_{i=1}^{n} X_{i}^{2}~, \qquad\mathrm{with} \qquad\langle
\beta\rangle= \beta_{0}~, \quad\langle\beta^{2} \rangle- \beta_{0}^{2}
=\frac{2}{n}\beta_{0}^{2}~.
\end{equation}
If the time scale over which $\beta$ fluctuates is much larger than the
typical time scale of order $\gamma^{-1}$ that the Langevin system
(\ref{langevinbeck}) need to reach equilibrium, then the conditional
probability $p(u|\beta)$ that the velocity takes on the value $u$, given a
value of $\beta$) is Gaussian and reads approximately:
\begin{equation}
p(u|\beta) = \sqrt{\frac{\beta}{2\pi}}\mathrm{exp}\left(  -\frac{1}{2}\beta
u^{2} \right)  ~.
\end{equation}
The probability to observe \emph{both} a certain value of $u$ and $\beta$
reads then:
\begin{equation}
\label{joint}p(u, \beta) = p(u|\beta)p(\beta)
\end{equation}
and therefore the probability of observing a value $u$ regardless of the value
of $\beta$ is:
\begin{equation}
p(u) = \int p(u|\beta)p(\beta) d\beta~,
\end{equation}
which can be evaluated~\cite{beck}, using (\ref{chidistribution}), to yield:
\begin{equation}
p(u) = \frac{\Gamma(\frac{n}{2} + 1)}{\Gamma(\frac{n}{2})}\left(  \frac
{\beta_{0}}{\pi n}\right)  ^{1/2} \frac{1}{( 1 + \frac{\beta_{0}}{n} u^{2}
)^{(n+1)/2}}~. \label{nesb}%
\end{equation}
Expression (\ref{nesb}) coincides (up to irrelevant proportionality constants)
with the generalised canonical distribution of non-extensive statistical
mechanics of Tsallis~\cite{tsallis}:
\begin{equation}
p(u) \propto\frac{1}{\left(  1 + \frac{{\tilde\beta}}{2}(q - 1)u^{2}\right)
^{1/(q-1)}}~,
\end{equation}
provided one makes the identifications $q = 1 + \frac{2}{n + 1}~,$ and
${\tilde\beta} = \frac{2}{3 - q}\beta_{0} ~$.

In \cite{beck} more general cases have been considered, where the Langevin
equation (\ref{langevinbeck}) is extended to include arbitrary frictional
forces of the form $F(u) = -\frac{\partial}{\partial u}V(u)$.


\begin{thebibliography}{99}
\bibitem {green} M. B. Green, J. H. Schwarz and E. Witten, \emph{Superstring theory}, Vols 1 \& 2
(Cambridge University Press, 1987).



\bibitem {polch}J.~Polchinski, \textit{String theory } Vols. 1 \& 2 (Cambridge University
Press, 1998).\
%\href{http://www.slac.stanford.edu/spires/find/hep/www?irn=4634802}{SPIRES entry}



\bibitem{sabmav} N.~E.~Mavromatos and S.~Sarkar,
  %``Towards a microscopic construction of flavour vacua from a space-time foam
  %model,''
  New J.\ Phys.\  {\bf 10}, 073009 (2008)
  [arXiv:0710.4541 [hep-th]].
  %%CITATION = NJOPF,10,073009;%%




\bibitem {Dfoam} J.~R.~Ellis, N.~E.~Mavromatos and D.~V.~Nanopoulos,
%``A microscopic recoil model for light-cone fluctuations in quantum
  %gravity,''
  Phys.\ Rev.\  D {\bf 61}, 027503 (2000)
  [arXiv:gr-qc/9906029];
  %``Dynamical formation of horizons in recoiling D-branes,''
  Phys.\ Rev.\  D {\bf 62}, 084019 (2000)
  [arXiv:gr-qc/0006004];
  %%CITATION = PHRVA,D62,084019;%%
%``Derivation of a Vacuum Refractive Index in a Stringy Space-Time Foam
  %Model,''
  Phys.\ Lett.\  B {\bf 665}, 412 (2008)
  [arXiv:0804.3566 [hep-th]].
  %%CITATION = PHLTA,B665,412;%%
J.~R.~Ellis,
N.~E.~Mavromatos and M.~Westmuckett,
%``A supersymmetric D-brane model of space-time foam,''
Phys.\ Rev.\ D \textbf{70}, 044036 (2004) [arXiv:gr-qc/0405066].
%%CITATION = PHRVA,D70,044036;%%


\bibitem {polch2}J. Polchinski, Phys. Rev. Lett. \textbf{75} 4724 (1995).


\bibitem {johnson}C. V.\ Johnson, \ \textit{D-Branes} (Cambridge University
Press, 2003).


\bibitem {streater} R. F. Streater and A. S. Wightman,\textit{\ PCT, Spin and
Statisitics, and All That}, (Benjamin, New York, 1964);
O.~W.~Greenberg,
  %``Why is CPT fundamental?,''
  Found.\ Phys.\  {\bf 36}, 1535 (2006)
  [arXiv:hep-ph/0309309].
  %%CITATION = FNDPA,36,1535;%%




\bibitem {bernabeu1} J.~Bernabeu, N.~E.~Mavromatos and J.~Papavassiliou,
  %``Novel type of CPT violation for correlated EPR states,''
  Phys.\ Rev.\ Lett.\  {\bf 92}, 131601 (2004)
  [arXiv:hep-ph/0310180];
  %%CITATION = PRLTA,92,131601;%%
J.~Bernabeu, J.~R.~Ellis, N.~E.~Mavromatos, D.~V.~Nanopoulos and J.~Papavassiliou,
\emph{CPT and quantum mechanics tests with kaons}, in A. Di Domenico, A. (ed.): \emph{Handbook on neutral kaon interferometry at a $\Phi$-factory}, 39 (2008)
[arXiv:hep-ph/0607322].
%%CITATION = HEP-PH/0607322;%%






\bibitem {bernabeu} J.~Bernabeu, N.~E.~Mavromatos and S.~Sarkar,
  %``Decoherence induced CPT violation and entangled neutral mesons,''
  Phys.\ Rev.\  D {\bf 74}, 045014 (2006)
  [arXiv:hep-th/0606137].
  %%CITATION = PHRVA,D74,045014;%%


\bibitem {wu}T.T. Wu and C. N. Yang, Phys. Rev.\ Lett. \textbf{13,} 380 (1964).

\bibitem {lipkin}  H.~J.~Lipkin,
  %``CP violation and coherent decays of kaon pairs,''
  Phys.\ Rev.\  {\bf 176}, 1715 (1968).
  %%CITATION = PHRVA,176,1715;%%


\bibitem {wheeler}J. A. Wheeler and K. Ford,\textit{\ Geons, Black Holes and
Quantum Foam}: \textit{A\ Life in Physics }(Norton, New York, 1998).

\bibitem {sarkar} S. Sarkar, J Phys A \textbf{41 }, 304013 \ (2008).

\bibitem {beck2}C. Beck and E.G.D. Cohen, Physica A \textbf{ 322 }, 267 (2003).

\bibitem {tsallis}C. Tsallis, J. Stat. Phys.\textbf{ 52 }, 479 (1988).

\bibitem {mavromatos} N.~E.~Mavromatos and S.~Sarkar,
  %``Liouville decoherence in a model of flavour oscillations in the  presence
  %of dark energy,''
  Phys.\ Rev.\  D {\bf 72}, 065016 (2005)
  [arXiv:hep-th/0506242].
  %%CITATION = PHRVA,D72,065016;%%


\bibitem {zwiebach}B. Zwiebach, \emph{A first course in string
theory} (Cambridge Univ. Press (2004)).

\bibitem {szabo} N.~E.~Mavromatos and R.~J.~Szabo,
  %``Matrix D-brane dynamics, logarithmic operators and quantization of
  %noncommutative spacetime,''
  Phys.\ Rev.\  D {\bf 59}, 104018 (1999)
  [arXiv:hep-th/9808124].
  %%CITATION = PHRVA,D59,104018;%%



\bibitem {coll} W. Fischler, S. Paban, and M. Rozali, Phys Lett \textbf{B381}, 62 (1996).

\bibitem {kogan} I.~I.~Kogan, N.~E.~Mavromatos and J.~F.~Wheater,
  %``D-brane recoil and logarithmic operators,''
  Phys.\ Lett.\  B {\bf 387}, 483 (1996)
  [arXiv:hep-th/9606102];
  %%CITATION = PHLTA,B387,483;%%
I.~I.~Kogan and N.~E.~Mavromatos,
  %``World-Sheet Logarithmic Operators and Target Space Symmetries in String
  %Theory,''
  Phys.\ Lett.\  B {\bf 375}, 111 (1996)
  [arXiv:hep-th/9512210].
  %%CITATION = PHLTA,B375,111;%%


\bibitem {mav2}  N.~E.~Mavromatos,
\emph{Logarithmic conformal field theories and
strings in changing backgrounds},
in \emph{Shifman, M. (ed.) et al.: From
fields to strings, I. Kogan memorial Volume 2}, 1257-1364. (World Sci. 2005),
and references therein. [arXiv:hep-th/0407026].
 %%CITATION = HEP-TH/0407026;%%

\bibitem {mathieu}P. Di Francesco, P. Mathieu, and D. S\'{e}n\'{e}chal,
\textit{Conformal Field Theory} (Springer 1997).

\bibitem {klebanov}I. Klebanov and L. Susskind, Phys. Lett. B\textbf{200} 446 (1988).

\bibitem {coleman}S. Coleman, Nucl. Phys. B\textbf{307} \ 867 (1988); Nucl.
Phys. B\textbf{310} 643 (1988).

\bibitem {zam}A.~B.~Zamolodchikov,
%``'Irreversibility' Of The Flux Of The Renormalization Group In A 2-D Field
%Theory,''
JETP Lett.\ \textbf{43}, 730 (1986) [Pisma Zh.\ Eksp.\ Teor.\ Fiz.\ \textbf{43}, 565
(1986)].
%%CITATION = JTPLA,43,730;%%

\bibitem {gardiner2}C. W. Gardiner, \emph{Handbook of Stochastic Methods}, 2nd Edn.,
(Springer 1990).

\bibitem {beck}C. Beck, Phys. Rev. Lett. \textbf{87} (2001) 180601-1.

\bibitem{fs} W. Fischler and L. Susskind, Phys. Lett. {\bf B171} (1986) 383;
{\bf B173} (1986) 262.

\bibitem {mavsarkar} N.~E.~Mavromatos, A.~Meregaglia, A.~Rubbia, A.~Sakharov and S.~Sarkar,
  %``Quantum-Gravity Decoherence Effects in Neutrino Oscillations: Expected
  %Constraints From CNGS and J-PARC,''
  Phys.\ Rev.\  D {\bf 77} (2008) 053014
  [arXiv:0801.0872 [hep-ph]].
  %%CITATION = PHRVA,D77,053014;%%

\bibitem{yoneya} T. Yoneya, Mod. Phys. Lett. {\bf A4} (1989) 1587.

\bibitem{liyoneya} M. Li and T. Yoneya, Phys. Rev. Lett. {\bf 78} (1997)
1219; {\it Short-distance Space--time Structure and Black Holes in String
Theory: A Short Review of the Present Status}, hep-th/9806240, to appear in
Chaos, Solitons and Fractals; \\ A. Jevicki and T. Yoneya, {\it Space--time
Uncertainty Principle and Conformal Symmetry in D-particle Dynamics},
hep-th/9805069, to appear in Nucl. Phys. {\bf B};\\ D. Minic, {\it On the
Space--time Uncertainty Principle and Holography}, hep-th/9808035.


\bibitem{emnuncertnew} J.~Ellis, N.~E.~Mavromatos and D.~V.~Nanopoulos,
  %``Derivation of a Vacuum Refractive Index in a Stringy Space-Time Foam
  %Model,''
  Phys.\ Lett.\  B {\bf 665} (2008) 412.
  %%CITATION = PHLTA,B665,412;%%

\bibitem{szabo2} N.~E.~Mavromatos and R.~J.~Szabo,
  %``The Neveu-Schwarz and Ramond algebras of logarithmic superconformal  field
  %theory,''
  JHEP {\bf 0301}, 041 (2003)
  [arXiv:hep-th/0207273];
  %%CITATION = JHEPA,0301,041;%%
  %``D-brane dynamics and logarithmic superconformal algebras,''
  JHEP {\bf 0110}, 027 (2001)
  [arXiv:hep-th/0106259].
  %%CITATION = JHEPA,0110,027;%%

\bibitem{dafne} F.~Ambrosino {\it et al.}  [KLOE Collaboration],
  %``First observation of quantum interference in the process Phi --> K(S) K(L)
  %--> pi+ pi- pi+ pi-: A test of quantum mechanics and CPT symmetry,''
  Phys.\ Lett.\ B {\bf 642}, 315 (2006)
  [arXiv:hep-ex/0607027];
  %%CITATION = HEP-EX 0607027;%%
   F.~Bossi, E.~De Lucia, J.~Lee-Franzini, S.~Miscetti and M.~Palutan  [KLOE Collaboration],
  %``Precision Kaon and Hadron Physics with KLOE,''
  Riv.\ Nuovo Cim.\  {\bf 031} (2008) 531
  [arXiv:0811.1929 [hep-ex]];
  %%CITATION = RNCIB,031,531;%%
  See also: A.~Di Domenico,
  %``Neutral kaon interferometry at Phi-factory,''
%\href{http://www.slac.stanford.edu/spires/find/hep/www?irn=7859015}{SPIRES entry}
{\it  In *Di Domenico, A. (ed.): Handbook of neutral kaon interferometry at a Phi-factory* 1-38} and
 and references therein.

\bibitem{lmn} See for instance: A.~B.~Lahanas, N.~E.~Mavromatos and D.~V.~Nanopoulos,
  %``Smoothly evolving Supercritical-String Dark Energy relaxes
  %Supersymmetric-Dark-Matter Constraints,''
  Phys.\ Lett.\  B {\bf 649}, 83 (2007)
  [arXiv:hep-ph/0612152];
  %%CITATION = PHLTA,B649,83;%%
B.~Dutta, A.~Gurrola, T.~Kamon, A.~Krislock, A.~B.~Lahanas, N.~E.~Mavromatos and D.~V.~Nanopoulos,
  %``Supersymmetry Signals of Supercritical String Cosmology at the Large Hadron
  %Collider,''
  arXiv:0808.1372 [hep-ph], Phys. Rev. D in press.
  %%CITATION = ARXIV:0808.1372;%%

\end{thebibliography}
\end{document}